# Fluorescence fluctuations-based super-resolution microscopy techniques: an experimental comparative study

Ida S. Opstad,[1, *] Sebastian Acuña,[1] Luís Enrique Villegas Hernandez,[1] Jennifer Cauzzo,[2] Nataša Škalko-Basnet,[2] Balpreet S. Ahluwalia[1], Krishna Agarwal[1,#]

[1]Department of Physics and Technology, and [2]Department of Pharmacy, UiT The Arctic University of Norway, NO-9037 Tromsø, Norway.

*ida.s.opstad@uit.no ; #uthkrishth@gmail.com

## Abstract

Fluorescence fluctuations-based super-resolution microscopy (FF-SRM) is an emerging field promising low-cost and live-cell compatible imaging beyond the resolution of conventional optical microscopy. A comprehensive overview on how the nature of fluctuations, label density, out-of-focus light, sub-cellular dynamics, and the sample itself influence the reconstruction in FF-SRM is crucial to design appropriate biological experiments. We have experimentally compared several of the recently developed FF-SRM techniques (namely ESI, bSOFI, SRRF, SACD, MUSICAL and HAWK) on widefield fluorescence image sequences of a diverse set of samples (namely liposomes, tissues, fixed and living cells), and on three-dimensional simulated data where the ground truth is available. The simulated microscopy data showed that the different techniques have different requirements for signal fluctuation to achieve their optimal performance. While different levels of signal fluctuations had little effect on the SRRF, ESI and SACD images, image reconstructions from both bSOFI and MUSICAL displayed a substantial improvement in their noise rejection, z-sectioning, and overall super-resolution capabilities.

## Abbreviations

(b)SOFI: (balanced) super-resolution optical fluorescence imaging
ESI: entropy-based super-resolution imaging
FF-SRM: fluorescence fluctuations-based super-resolution microscopy
HAWK: Haar wavelet kernel
MUSICAL: multiple signal classification algorithm
TIRFM: total internal reflection fluorescence microscopy
SACD: super-resolution imaging with autocorrelation two-step deconvolution
SBR: signal-to-background ratio
SIM: structured illumination microscopy
SNR: signal-to-noise ratio
SMLM: single molecule localization microscopy
SRM: super-resolution microscopy
SRRF: super-resolution radial fluctuations SRM: Super-resolution microscopy
STED: stimulated emission depletion microscopy
2D/3D: two/three-dimensional

## Introduction

Super-resolution microscopy (SRM) has revolutionized the field of microscopy, allowing visualization of nanoscale sub-cellular details smaller than the diffraction limit of optical microscopy. The spectrum of techniques in SRM is spanned by single molecule localization microscopy (SMLM), stimulated emission depletion microscopy (STED) and structured illumination microscopy (SIM). All SRM techniques require an expensive high-end acquisition system, expert sample preparation and system operation. Live-cell imaging is demonstrated for all of these SRM techniques [1], but remains extremely challenging because of especially two reasons. Firstly, the fast dynamics of many cellular processes in combination with relatively weak fluorescent signal, render acquisition of sufficient signal-to-noise ratio (SNR) for most analytical tasks challenging. Secondly, the cellular functions and morphology are sensitive to small changes in the cellular biochemical environment that can be significantly altered by introducing fluorescent probes, imaging buffers and excitation light exposure. As a consequence, SIM is arguably the best SRM technique for living samples currently available due to its comparatively fast widefield and volumetric acquisition together with lesser requirements on fluorophore photophysical properties and illumination intensities. However, under sub-optimal acquisition conditions such as fast-moving samples, low signal-to-background ratio (SBR) and/or significant photobleaching, SIM reconstruction often fails and is prone to reconstruction artifacts. Furthermore, the SIM imaging systems are not commonly available, likely due to their cost and complexity, and the requirement for trained personnel for system maintenance and operation.

| Structured illumination microscopy | Fluorescence Fluctuation based Super-Resolution Microscopy (FF - SRM) | Localization microscopy and STED |
|---|---|---|
| Requires costly equipment | No special requirements on instruments or dyes | May need specialized equipment |
| Needs photostable labels | Uses natural photophysics of fluorescence | Special blinking photophysics |
| Few images at high intensity | Camera-speed imaging rates | Poor live-cell compatibility |
| Better live-cell compatibility | Live-cell compatible imaging environment | Long imaging time |
| Resolution 120-160 nm | Resolution 50 – 120 nm (technique dependent) | Resolution 20-50 nm |

| SOFI | MUSICAL | ESI | SACD | SRRF |
|---|---|---|---|---|
| √ Quantitative (bSOFI)<br>√ Well-understood<br>√ High-resolution possible<br>× Sensitive to noise<br>× Many frames needed<br>× Requires stationarity<br>× Prone to artefacts | √ Good z-sectioning<br>√ Preserves motion<br>√ High-resolution possible<br>√ Few to several frames<br>× Intensity is blinking dependent<br>× Prone to artefacts | √ Noise removal<br>√ Few frames needed<br>× Poor resolution enhancement<br>× Overly bright and large spots where higher emitter densities | √ Few frames needed<br>√ Good for 3D samples and live-cells<br>× Poor resolution enhancement<br>× Over-slimming artifact | √ Few frames needed<br>× Poor z-sectioning (TIRF/2D only)<br>× Many option selections needed<br>× Requires stationarity<br>× Fits fibrous structures |
| Sensitive to noise – create mesh like artefacts<br>Benefit from high levels of fluctuations<br>Benefit from higher frame number | | Relatively better at noise suppression irrespective of noise level<br>Relatively insensitive to levels of fluctuations<br>Do not significantly benefit from higher frame number | | |
| • Use bSOFI option<br>• Use labels with strong fluctuations<br>• Use for >1000 frames & stationary samples | • Use for mobile structures or labels with strong fluctuations<br>• Use for 3D or 2D samples | • Use for noise suppression in samples of low intensity dynamic range | • Use for tissue samples for contrast enhancement<br>• Use when few frames are available | • Use for fibrous and 2D structures |

*Figure 1: Summary of our observations and recommendations for FF-SRM and comparison to other super-resolution microscopy techniques.*

There is a new set of techniques, namely fluorescence fluctuations-based super-resolution microscopy (FF-SRM) techniques that, like SMLM, use the photokinetics of fluorescence emission, but do not rely on the external introduction of spatio-temporal sparsity via the chemical environment and high-power laser modulation. This is an interesting avenue for bio-image analysis, possibly with the potential of democratizing SRM by greatly reduced system cost, and overall live-cell capabilities of high-resolution microscopy. The core phenomenon utilized in FF-SRM is the stochasticity of the number of photons emitted by fluorescent labels over time. These techniques use statistical analysis as the core mechanism to super-resolve the fluorescent molecule distribution, where each molecule independently contributes to fluctuations in the measured fluorescence intensity. FF-SRM in the context of super-resolution fluorescence microscopy techniques is presented in Figure 1.

Although the development of FF-SRM techniques is fairly recent, several techniques have been proposed in the short duration of a few years. Each of these techniques differs in the treatment of the raw data and statistical approach used. Some of them are super-resolution optical fluorescence imaging (SOFI) [2] and balanced SOFI (bSOFI) [3], entropy-based super-resolution imaging (ESI) [4], super-resolution radial fluctuations (SRRF) [5], multiple signal classification algorithm (MUSICAL) [6], super-resolution imaging with autocorrelation two-step deconvolution (SACD) [7], Bayesian analysis of blinking and bleaching (3B) [8], and sparsity based super-resolution correlation optical microscopy (SPARCOM) [9]. Additionally, the data pre-processing technique Haar wavelet kernel (HAWK) analysis has been developed as a tool to enable SRM of higher-density emitter data for both SMLM and FF-SRM, thus 'enabling high-speed, artifact-free super-resolution imaging of live cells' [10].

As evaluated and benchmarked in the original papers (by using reference examples from single molecule localization microscopy dataset and simulation examples), they provide a resolution in the range of 50-120 nm. Notably, all of the above-mentioned FF-SRM techniques use two-dimensional (2D) PSF considerations only (not 3D), and the simulated emitters lie perfectly in the focal plane, except for the noteworthy exception shown by Solomon et al. [9], where also emitters at 1 μm distance from the focal plane were considered. More details on the individual methods and their reconstruction parameters are provided in the Supplementary Methods.

When imaging real three-dimensional samples for biological or biomedical applications, the reliability of the reconstruction is of more significance than any of the quantitative merits such as the image resolution or contrast. We are not aware of any comprehensive study of how these methods perform on real biological samples in comparison to each other and under various conditions of intensity fluctuation.

Each of the methods has been demonstrated on experimental data of samples that have been arguably designed to illustrate the best characteristics of their own method or on SMLM benchmark data in which case all the methods benefit from the spatio-temporal sparsity in the fluorescence. A comparative study of these techniques on a wide variety of data is important to understand the opportunities and potential pitfalls of the different methods. Therefore, an in-depth analysis is needed on the sample and imaging conditions and how they affect the performances of FF-SRM methods. For example, how the sample and label density, out-of-focus signal, nature of fluctuations, and sub-cellular dynamics affect the reconstruction would be insightful for the experimental design and choice of technique. Moreover, such a comparative study will contribute in setting the right expectations and assigning suitable confidence in the biological interpretations derived from these methods. To this end, we have undertaken a first large-scale experimental study of FF-SRM techniques covering the following aspects:

1. We present an extensive study encompassing nanoparticles (liposomes), actin and membrane in fixed cells and tissues, and mitochondria and the endoplasmic reticulum (ER) in living cells.
2. We tested all the methods on exactly the same data, thereby performing the first unbiased comparative analysis of the performances of the techniques. In most situations, factors such as fluctuation density, number of frames, and a variety of relevant conditions for imaging or processing the data are considered. The control parameters of each method are tuned within reasonable limits to identify the best performance of the method and the related algorithmic settings.

3. We elucidate the performances of the techniques through three-dimensional (3D) simulation examples that closely emulate the sample conditions. We explain how and why the actual samples challenge the fluctuations-based techniques beyond the scope of design. We consider effects such as out-of-focus light, density of labeling, temporal density of photon emission, practical noise models and the number of frames used for reconstruction.

4. We elucidate the favorable conditions for methods and highlight the challenges that must be addressed in the algorithmic development of these FF-SRM techniques towards making them reliable tools in biomedical research.

## Results and Discussion

### Simulated data

To obtain fair and definitive answers about the different methods' performance, simulated samples with known ground truth were generated. Two different 3D test samples with varying levels of intensity fluctuations were generated and processed using ESI, SRRF, SACD, SOFI and MUSICAL. The ground truth emitter locations with axial color coding as a distance from the focal plane are displayed in the upper panels of Figure 2, while their corresponding microscopy images are displayed in the panels below (simulating 510 nm emission wavelength and 1.42NA microscope objective). The biological relevance and structural details of the two samples are as follows:

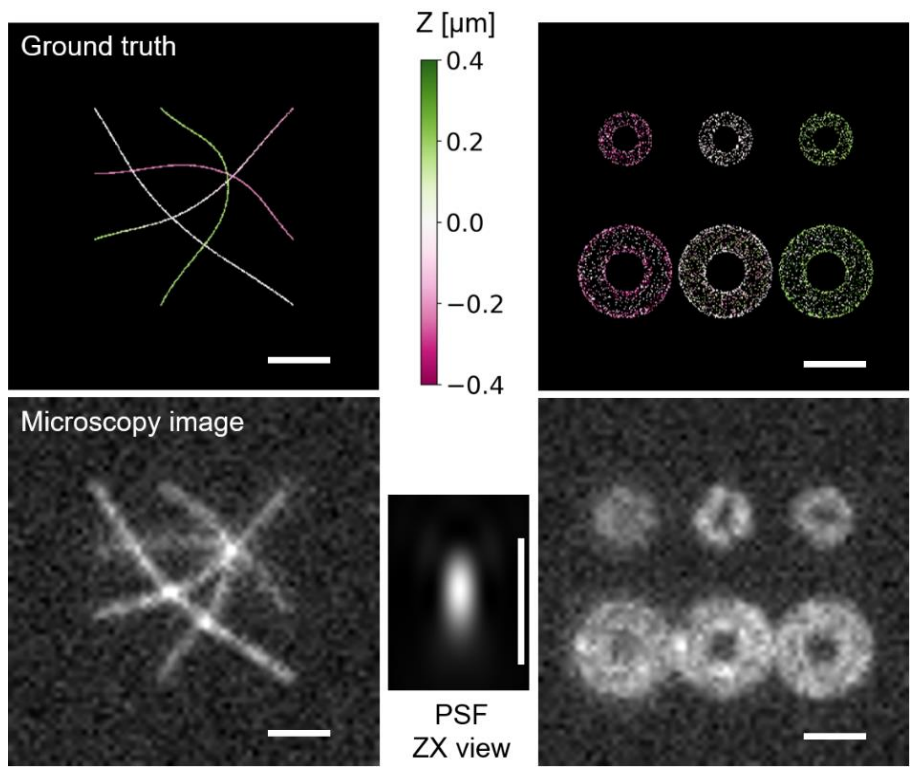

*Figure 2: The top row displays the simulated data's ground truth with color coded z-position compared to the focal plane (Z=0). The bottom row shows the corresponding microscopy images (single frames) after noise addition and the simulated PSF (orthogonal view) using 1.42NA and emission wavelength 510 nm. The Abbe resolution limits under these conditions are laterally 180 nm and axially 506 nm. The scale bars are 1 μm.*

1) **Actin strands**. There are four non-intersecting actin strands. Three strands are parallel with the coverglass in different z-planes, one in the focal plane and two of them above and below the focal plane by 0.4 μm respectively. The fourth actin strand is inclined and positioned across several z-planes, having one end 0.2 μm below and the other 0.2 μm above the focal plane. None of the strands are physically touching, but their (projected) microscopy image has overlapping signal in the regions where their lateral positioning is the same. These overlapping regions are where the algorithms' performance is of particular interest.

2) **Tori (hollow doughnuts).** The upper row of tori corresponds to tubes of 200 nm diameter, while the lower row has tubes of 400 nm diameter. Both rows have tori centered at three different z-positions. The tori in the lower row are resolvable using conventional microscopy, while the tori in the upper row are not. These structures were chosen to emulate significant cellular organelles like mitochondria and the endoplasmic reticulum (ER) which are outlined by 3D tubular membranes. To resolve both the inner and outer peripheries, the FF-SRM methods must exhibit a good z-sectioning, recognition of small intensity differences but only minor lateral resolution improvement compared to the diffraction limit of optical microscopy.

A higher number of frames for the reconstructions (5000 frames) were used for ESI and SOFI compared to the other techniques (16 to 100 frames). This was due to negative results of initial testing, their capability of fast computations for larger stack sizes, together with the much higher frame number indicated by the methods' original publications.

We will especially consider three aspects of the reconstructions: i) background signal and effect of noise, ii) reconstruction quality and artifacts, iii) the effect of out-of-focus objects and z-sectioning abilities.

**Actin strand simulations**

The best results achieved from a variation of tested parameters by the five FF-SRM methods are displayed in Figure 3 in the case of simulated 3D actin strands for different levels of intensity fluctuations. A higher level of intensity fluctuations was achieved via sparser fluorescence emission from individual molecules on a densely labelled sample. The different levels are defined quantitatively in the supplementary information.

The *noise* present in the simulated microscopy images (Figure 2) appears not to pose a challenge to ESI, SRRF or SACD. The *structural representations* are accurate except at the intersections of the actin strands (or their projected images). Specifically, in the case of ESI the joints are excessively large and bright (the images are non-linearly intensity adjusted to also allow for visualization of the dimmer structures), and in the case of SRRF and SACD, the strands are completely missing close to the intersections. The performance of ESI, SRRF and SACD appears also largely unaffected by the varying level of fluorescence fluctuations, except for an additional out-of-focus strand appearing in the ESI images at higher levels of fluorescence fluctuations. SRRF does not exclude out-of-focus signal, while SACD does, both independently of the level of intensity fluctuations.

This is very different from the results of SOFI and MUSICAL; whose performance was highly dependent on signal fluctuation level. As opposed to ESI, MUSICAL rejects more out-of-focus structures the higher the level of intensity fluctuations, and the reconstruction of the in-focus sample area are notably better. SOFI and MUSICAL do not appear to have the same issues close to the intersection points as ESI, SRRF and SACD, but SOFI is badly affected by the noise, which results in a dominating background signal that could be difficult to distinguish from the image objects. HAWK preprocessing alleviated the background issue of SOFI, especially for the highest level of fluorescence fluctuations. No improvement was found using HAWK for the other techniques. Further results using additional reconstruction parameters and other image stack sizes are found in Suppl. Figure S1 together with a more elaborate discussion on the performance of the different techniques and their artifacts under varying conditions.

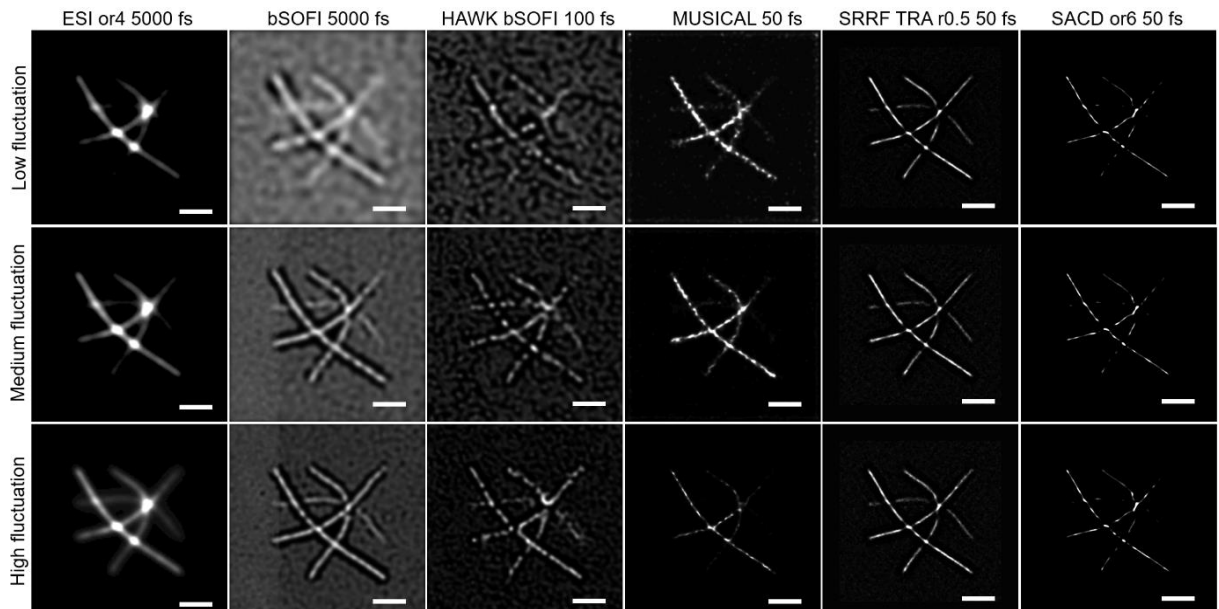

Figure 3: FF-SRM reconstructions of simulated actin strands for three different levels of fluctuations for all five tested methods. Note that only one of the strands lies completely in the focal plane. The bSOFI and MUSICAL images are clearly improved for higher fluctuation levels, while the ESI, SRRF and SACD images display no improvement for higher levels of fluctuations. The headers indicate method and some details about the reconstruction parameters: fs: number of frames; or: order; TRA r0.5: temporal radiality average with SRRF ring radius 0.5. The ESI images are intensity adjusted using γ = 0.5 intensity adjusted, while all other panels have linearly adjusted intensities. The scale bars are 1 μm.

**Mitochondria/tori simulations**

Although useful insights can be derived from simple examples like crossing actin strands, they are too simplistic to reveal how the techniques might perform on more complex biological structures such as 3D tubes.

The results for the simulated tori are summarized in Figure 4 for two different fluctuation levels and for each case one torus centered at perfect focus and one 200 nm above the focal plane. These tori correspond to the upper right and middle torus of Figure 2. Results for the complete sample are available in the SI together with results using additional reconstruction parameters (Suppl. Figures S2-S4).

As also noted for the actin strand example, ESI, SRRF and SACD eliminate noise and appear insensitive to fluctuation level as well as the 200 nm shift from the focal plane. Compared to the ground truth structures, which no longer are single lines, none of these techniques can make out the double rings (or 3D tubes). SRRF and SACD reconstruct rings way too slim compared to the actual structures. This reconstruction artifact would not be noticeable using the actin strand example alone.

SOFI, as for the simulated actin strands, is sensitive to noise which gives some artifacts in the background but is able to reconstruct the tubes for the case of 5000 frames and a high fluctuation level (but fails for 100 frames or low level of fluctuations). For a high level of signal fluctuations, MUSICAL is able to discern the double ring of the in-focus torus for only 100 frames, but better for 1000 frames and then also for the low fluctuation level. MUSICAL does not show any background artifacts from the noise for these cases.

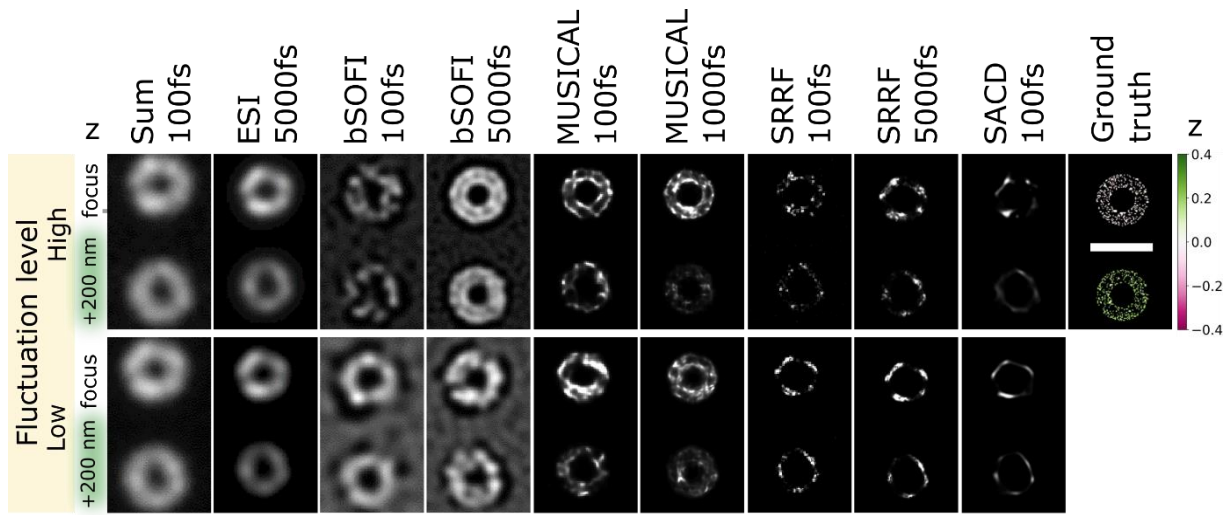

*Figure 4: : Reconstructions of tori (tubes of diameter 200 nm) for high (top row) and low signal fluctuation (bottom row), for a torus centered at the focal plane and 200 nm above focus. The tube shape emulates cellular structures like mitochondria and the ER. The scale bar in the ground truth image is 1 μm, and the color bar describes the emitters' axial positions in μm. Only MUSICAL manages to resolve the outer rings for 100 frames (in-focus torus at high fluctuation level), while SOFI provided good reconstruction using 5000 frames, but only for the high fluctuation level and still with significant background artifacts, likely cause by the simulated noise addition. Using 1000 frames, MUSICAL could resolve parts of the inner and outer circles also for the low fluctuation level. The ESI (γ = 0.5 intensity adjusted), SRRF and SACD results show only a single circle for each torus (for any number of frames or parameters tested), but also with complete noise removal. The circles are in the case of SRRF and SACD significantly slimmer than the ground truth 'double circle', which illustrates a typical reconstruction artifact with these techniques that can be difficult to spot when the ground truth is not available.*

These simulation examples have revealed some strengths and weaknesses with all five FF-SRM techniques under scrutiny. We will in the following consider their performance on actual experimental data and see how the results compare to the ones from the simulated data.

**Liposomes**

The small size, agile and delicate nature of liposomes make their characterization by microscopy challenging and a non-standard procedure. We tested three different sample preparations for liposomes with integrated fluorescence (NBD with excitation and emission maxima 476 nm and 537) directly on microscopy cover glasses: Free floating in suspension, dried-on, and small droplets immobilized under a patch of solid agarose gel.

The samples were imaged in fast time-lapse mode using standard epi-fluorescence microscopy. The free-floating liposomes were, as expected, moving too fast in especially axial direction for acquisition of multiple time point videos of the particles. The dried liposome suspensions appeared to be destroyed, while the suspensions of liposomes covered by solid agarose appeared intact and stationary over the course of 200-300 time points. Hence, only the samples with liposomes immobilized via agarose were considered for further analysis.

We tested the five FF-SRM methods' ability to accurately reveal liposome size from two different known size distributions: 100 nm and 250 nm, respectively. To this end, we first assessed the optimal number of frames to be used for the analysis (Suppl. Figure S5-S6). When not clear which number of frames were best, 100 frames were used, which in most cases was found to provide the optimal tradeoff between fluctuation data (i.e. number of frames) and (rapid) photobleaching together with potential instability of the supporting agarose. The autofluorescence of the agarose patch was also found to photobleach faster than the fluorophores for the first 100-200 frames, possibly beneficial to some of the FF-SRM methods.

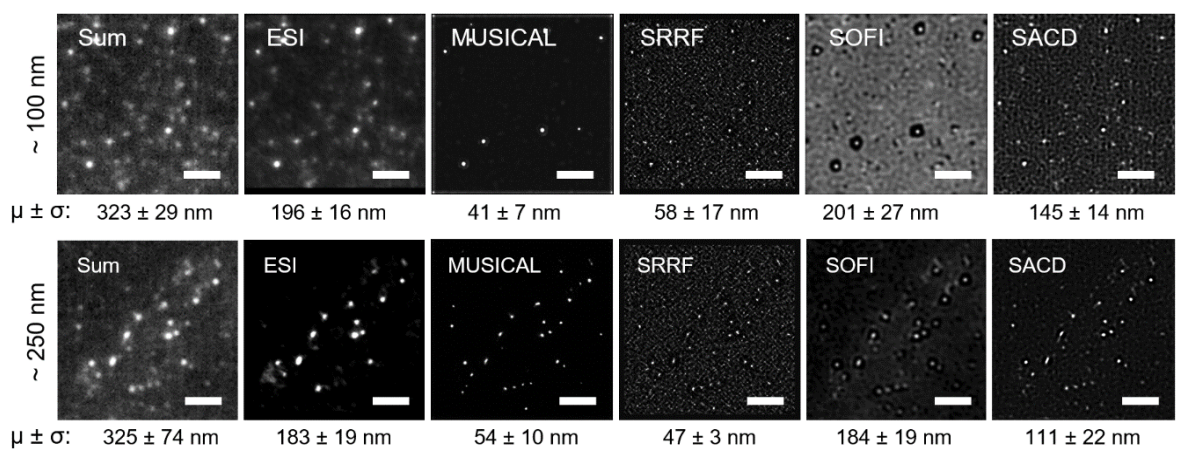

Figure 5: Reconstruction results for liposomes of about 100 nm (upper row) and 250 nm (bottom row) size distributions for the five different methods: ESI order 4 (100 fs), MUSCAL (100 fs) threshold -0.21 (100 nm) and -0.57 (250 nm), SRRF TRAC ring radius 0.5 (100 fs), bSOFI (100 fs for 100 nm and 200 fs for 250 nm), and SACD order 2 (100 fs for 100 nm and 25 fs for 250 nm). The mean value and standard deviation from measuring liposome FWHMs (Gaussian fit) are stated below the panels. Notably, the measured sizes depend on FF-SRM method (and their parameters) and seemingly not on the liposome size distribution. When 100 fs were used for both size distributions for SACD and SOFI, the mean values were 145 nm and 146 nm for SACD, and 201 nm and 203 nm for bSOFI.

Figure 5 shows the results evaluated as best for both the 100 and 250 nm liposomes for the five FF-SRM methods (additional results were available in Suppl. Figures S5-S7). From these images, five FWHM measurements for each case were measured, with the resulting mean and standard deviation displayed under the panels of the respective reconstructions. Notably, the estimated size depends on the FF-SRM method used, and seemingly not on the underlying liposome size distribution. Each technique gives a different result, but the same technique gives a similar result (< 35 nm difference on the mean value) for the two significantly different size distributions (about 100 nm and 250 nm). When the same number of frames were used for the two size distributions for SOFI and SACD (different #fs were found best for the two different size distribution for these cases), the difference was even smaller (1 nm for SACD and 2 nm for bSOFI), see captions of Figure 5 and Suppl. Figure S5. The individual measurements and chosen liposomes are shown in Suppl. Figure S8.

This small ensemble study illustrates some of the challenges with these FF-SRM methods. Although we cannot completely exclude the possibility that one of these techniques provides the right answer for all measured lipid particles (as the ground truth is not available), the size measurements seem completely off and unlikely to be correct for either technique. Changing any reconstruction parameters of the individual techniques also changed the measurements. For example, on the ~250 nm sample, using 25 frames for SACD gave 111 nm mean value for the FWHM, while using 100 fs resulted in mean of 146 nm. Similarly, SOFI with 100 fs gave 203 nm, while using 200 fs gave 184 nm mean value for the FWHM measurements. Better signal of the larger liposomes also appears to 'make the localization better' resulting in smaller size estimates (for all methods except MUSICAL, although also these size estimates are also clearly too small).

The agarose patch appears to have caused notably background artifacts in the reconstruction for SRRF, SOFI and SACD, but not as significantly for ESI or MUSICAL for these particular samples. This problem would likely be alleviated if a more stable fluorophore were available. This was however not the case for this sample, as fluorescent molecules in general are challenging to stably incorporate into liposomes.

The achieved image resolutions were estimated via line profiles over a sample area with an elongated spot, indicating the presence of at least two closely separated liposomes (Suppl. Figure S9). The MUSICAL, SRRF and SACD images show clear dips between two (or more) peaks, but the high prevalence of reconstruction artifacts in especially the SRRF and SACD images (likely caused by the agarose autofluorescence) render these measurements unreliable.

For future experiments, it might be of interest to ensure that the liposomes are arranged as a flat, monolayer sample that remains stably in perfect focus during image acquisition. Even small deviations from focus could alter the liposome size measurements. The use of total internal reflection fluorescence microscopy (TIRFM) would also likely help reducing the effects of agarose unevenness, autofluorescence and out-of-focus signal. These points could also be used as a general consideration for size profiling applications that use FF-SRM for particles of dimensions close to or smaller than the resolution limit.

Although quantitative analysis does not seem promising from this initial approach, it might be possible via calibration of the individual techniques' parameters on known size distributions to obtain more reliable size estimates. Especially the integration of more photostable fluorophores into the liposomes would be a game changer. As we saw from the simulation examples for the SOFI images, reliable reconstruction was not achieved for ~100 frames, but for 5000 frames with a high level of intensity fluctuations.

We will now proceed to samples where often the *qualitative* information is of considerable interest, namely biological tissues and cells.

**Fixed cells and tissues**

The five different FF-SRM methods were tested on fixed cell cultures (macrophages) and tissues (placenta and heart cryo-sections) using the commonly applied fluorescent probes CellMask Orange (membrane marker) and Phalloidin-ATTO647N (labeling filamentous actin), as before, illuminated using incoherent wide-field illumination for standard epi-fluorescence microscopy.

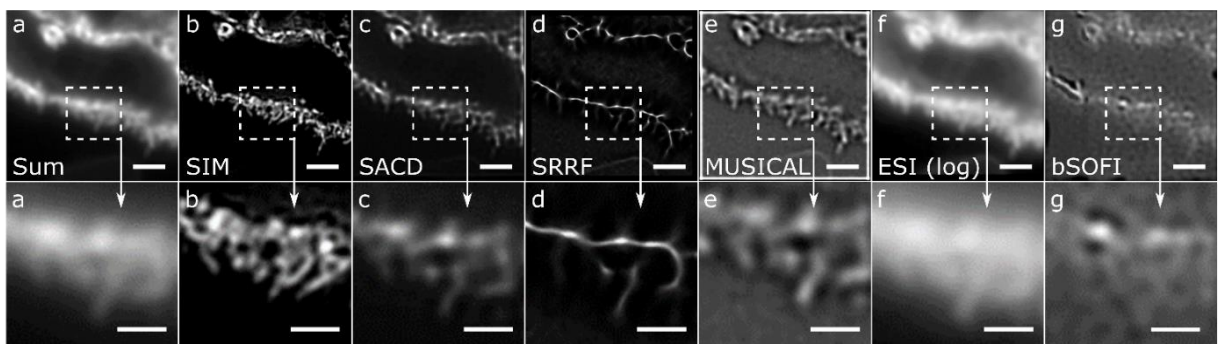

*Figure 6:  FF-SRM reconstructions of 1 μm-thick cryo-preserved placental tissue section fluorescently labelled with Phalloidin-ATTO647N for identification of F-actin. The regions indicated in the upper panels are shown magnified below revealing the microvilli brush-border of a chorionic villus. (a) The summed image of 500 frames; (b) a single z-plane 3D SIM image; (c) SACD using 50 frames and order 2; (d) SRRF using 500 frames along with TRA option and radius 0.5; (e) MUSICAL using 200 frames and threshold -0.33267; (f) ESI order 4 using 500 frames (log intensity adjusted); bSOFI using 500 frames. The scale bars are 2 μm on the upper-row panels and 1 μm in the lower-row panels.*

The results were evaluated from a broad range of different reconstruction parameters for the different methods and the results considered best for each method are displayed in Figure 6 and Suppl. Figure S10 for the case of placenta tissue, and Suppl. Fig S12 for fixed cells. Results using additional reconstruction parameters/options and a data overview are available in Suppl. Figures S11-S15.

The results for the different FF-SRM methods applied to the same sample are strikingly different. Comparing with the sum and the 'reference' SIM image (providing resolution doubling compared to the diffraction limit) of Figure 6, only SACD and MUSICAL give a minor improvement in detail visibility over conventional microscopy. The ESI image appears similar to the sum image, the SRRF image generates thin lines partly corresponding to the SIM image, while the SOFI image is a complete mesh of artifacts.

Results on ultrathin tissue sections (100 nm thickness) and TIRFM data gave similar discouraging results (Suppl. Figures S16-S18). This strongly indicates that out-of-focus signal is not the main reason for the methods' failure.

Comparing with the simulation results presented earlier, the results indicate that the high background intensities and in general poor performance of both MUSICAL and SOFI could be explained by the photo-physical properties of the fluorescent labels used, and that these problems could be countered by experimentally introducing a higher level of fluorescence intensity fluctuation (e.g. using different fluorophores or imaging conditions). Also using longer sequences (>400 frames) might have improved the results, this data is however not available.

**Living cells and dynamics**
One major motivation for performing FF-SRM instead of other nanoscopy techniques is the opportunity for data acquisition under live-cell friendly environment. In this section, we consider epi-fluorescence time-lapse data of living cells. Because of the dynamic and delicate nature of living cells, fewer frames and lower illumination intensities were used for these data sequences.

The different FF-SRM methods were applied to three different test samples: mitochondrial outer membrane and ER where little to no dynamics were visible in the conventional image stack (64 frames), and a 100 frames image sequence of mitochondria undergoing fast dynamics. The results on mitochondria for stationary and fast dynamics are displayed in Figure 7, while the results for ER and additional HAWK results for mitochondria are displayed in Suppl. Figure S19.

As seen for the fixed samples, all the different methods gave vastly different pictures when applied to the same image sequence. For the stationary sample, the reconstructions show similar patterns as seen for the fixed cells and tissues: ESI provides noise removal and structure slimming, but no real resolution improvement. MUSICAL provides a dominating artefact network over the entire object area. SRRF fits thin single lines to the wider tubular structure. SACD impresses with sturdily recognizing and reconstructing the outer mitochondrial membrane. The great improvement over the simulation results on the tori seen in Figure 3, can be explained by the real mitochondria (in this particular sample) are wider (~250-500nm) than the 200 nm tubes of the tori, and not beyond the resolution limit of SACD. This can be also seen from Suppl. Figure S3, where SACD results on the entire tori simulation sample is shown. Here, the SACD images of the larger tori (400 nm tubes) show two concentric circles, while the smaller tubes (200 nm) are represented as thin mono-circles. Notably, the mitochondrial outer membrane is discernible in some places in the raw data, and especially for the summed image.

HAWK preprocessing resulted in an overall noisy and degraded image, but also a more discernible outer membrane in the case of ESI, MUSICAL and SRRF. The ER sample displayed similar patterns of reconstruction artifacts as for the mitochondria but is also an extremely difficult sample to evaluate as this tubular membrane network could take on almost any shape (shown in Suppl. Figure S19).

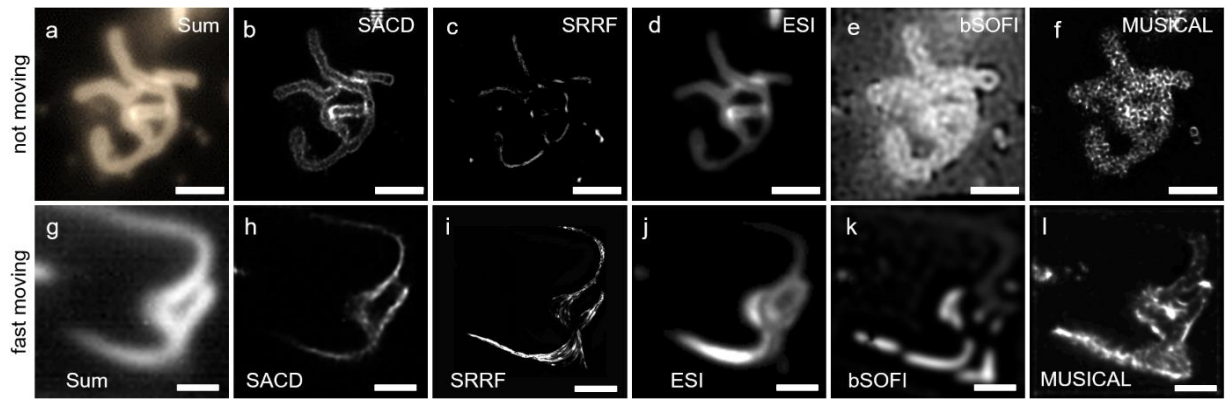

*Figure 7: Reconstructions on live-cell data of mitochondrial outer membrane (OMP25- mCherry). Top row: stationary organelles (scale bars: 2 μm); bottom row: fast moving mitochondria (scale bars: 1 μm). The mitochondrial dynamics introduce a new type of signal fluctuation that is not accounted for by any of the FF-SRM algorithms and introduces different artifacts compared to those of stationary objects. The object dynamics has a clearly different effect on all the five different methods. Interestingly, the mitochondrial outer membrane appears much better reconstructed by MUSICAL in the case of dynamic mitochondria compared to the stationary mitochondria.*

For the extremely dynamic sample, ESI appears similar to the sum image, SACD similar to a strongly deconvolved sum image, while SOFI has deleted parts of the moving structure, presumably because dynamics give less pixel-wise signal correlation. SRRF appears to fit a different thin line for every time-point, resulting in a fine grid of multiple lines. The MUSICAL image of mitochondria looks strikingly different from the one in the previous figure, with sharp contours of the outer membrane instead of the dominating artefact network seen in the previous figure and for the results on fixed cells. The signal fluctuations introduced by the mitochondrial dynamics appear to be exploited by the MUSICAL algorithm.

## Results and discussion summary

We have processed datasets from a broad range of samples and applied to them the fluctuation nanoscopy techniques (b)SOFI, ESI, MUSICAL SRRF and SACD, trying out many different reconstruction parameters along the way. Figure 1 presented a summary of our observations, which are discussed in detail below.

**Observations regarding SOFI and MUSICAL:** The simulations revealed that only two of the techniques, namely SOFI and MUSICAL, required a high level of intensity fluctuations to achieve their optimal results. Also only these techniques were able to resolve the more challenging 3D tube-like structures of 200 nm diameter, simulating membrane-bound cellular organelles like mitochondria and the ER. The SOFI images displayed dominating artifacts in presence of noise, but for data of high level of fluorescence signal fluctuations and thousands of time-point image sequences displayed reliable reconstruction even for the 3D samples. HAWK lowered SOFI's sensitivity to noise and greatly improved the SOFI reconstructions for few raw image (~100), but only for a high level of fluorescence intensity fluctuations. In the case of short image sequences with a high level of intensity fluctuations, MUSICAL performed the best. MUSICAL also showed an additional ability to exploit signal fluctuations arising from sample dynamics. For fixed cells and tissues, the disappointing performance of SOFI and MUSICAL was shown to be due to a too low level of signal fluctuations in our experimental data. This was especially inferred from the results on simulated data, where SOFI and MUSICAL displayed poor performance for low fluctuation levels, but good performance for higher fluctuation levels.

Additionally, SOFI and MUSICAL performed poorly in the case of slow-moving (or stationary) structures in living cells, both producing a dominating circular mesh. This can be explained by a low level of intensity fluctuations, but importantly, also the use of short image sequences (to assure sample stationarity). However, in the case of the fast-moving sample, MUSICAL was able to exploit the fluctuations induced by the sample dynamics, producing a significantly better results than seen for the slow-moving structures.

**Observations regarding ESI, SRRF and SACD:** Although ESI displayed faithful noise removal and was possibly the technique the least prone to artifacts, it failed to show super-resolution capabilities for our data. SRRF had also strong noise-reduction capabilities for all fluctuation levels but failed to reveal the true underlying structures where the ground truth (beyond the diffraction limit) was available. Both SRRF and SACD were shown to produce 'over-slimming' of structures, rather than revealing the true nanoscopic details in the case of the 3D simulations of doughnuts. In the case of low signal fluctuations and 'ultra-short' image sequences (16 frames), SACD had the decidedly best performance of 3D structures close to the resolution limit (like the mitochondrial outer membrane), although its tendency towards producing over-slimming artifacts must be kept in mind while analyzing SACD imaging results. We noted that for fixed cells and tissues, the performance of ESI, SRRF and SACD are generally better than for SOFI and MUSICAL in the sense that the images overall look closer to the actual samples with less obvious artifacts, even though they did not display super-resolving abilities. This is in agreement with our simulated 3D examples in the case of low fluctuations, where we did notice better robustness of these techniques irrespective of the super-resolution ability. Nonetheless, these techniques might generate subtle artefacts that are difficult to spot. The possible influence of these subtle artifacts in the analysis of bio-images needs further investigation. SACD showed a strong ability in producing reliable reconstructions for structural details close to the diffraction limit, as evident from the live-cell data of slow-moving mitochondria. None of them however could withstand the challenge of fast-moving mitochondria.

**General observations that apply to all the FF-SRM techniques under scrutiny:** The simulated 3D examples do provide some important insights into the performance of these methods. A significant one is that FF-SRM methods can perform well for actin or other fiber-like structures and these might be good examples for studying resolution. However, these results may not be suitable for setting the expectations regarding the performance of these methods for more complex 3D samples such as mitochondria and the ER. Two more important insights from simulations are regarding (a) the effect of out-of-focus structures and level of fluctuations on the reconstructions and (b) the artefacts arising from noise and overlapping structures.

Our results showed an overall poor performance of all FF-SRM methods for the tested conditions for liposomes, fixed cells and tissues. We noted that even if the samples are ultrathin or optical sectioning is not a challenge, FF-SRM can often fail in the case of low fluctuation levels, high background signal and/or insufficient data (number of frames). The measured sizes of liposomes from different known size distributions, revealed that the measured FWHM depend more on chosen FF-SRM technique than on nanoparticle size. Further experimental optimization and calibration of the individual methods reconstruction parameters would be needed before trustworthy nanoparticle size measurements can be carried out using FF-SRM.

The use of dense labelling and photo-stable fluorophores that are optimal for other nanoscopy techniques led to failed reconstructions and image artifacts in the case of fixed cells and tissues. Nonetheless, acquiring a large number of frames, using better-suited dyes, and introducing a higher level of fluctuations through use of imaging buffers, might assist these techniques in performing better. Depending on the resolution requirements and system availability, it might be preferable to use SRM techniques like SIM, STED or localization microscopy for fixed cells and tissues, as the considerations related to live-cell imaging do not apply for these samples.

The many different parameters offered by some of the techniques could be a potential strength allowing for super-resolution imaging for a broader range of samples and imaging conditions. It is however problematic that, to our knowledge, there are no clear guidelines for when the different parameters should be used, leaving the user with difficult and subjective choices about what might be 'the best' reconstruction. Usually the ground truth is not available for bio-image data, which only complicates the path to derive good guidelines for parameter selection.

Live-cell compatibility is advertised by all evaluated FF-SRM methods. Still, and somewhat unfortunately, stationarity of the imaged objects (during the course of the analyzed image sequence) is also assumed by the FF-SRM algorithms (all apart from MUSICAL). Our computational experiments on extremely dynamic samples displayed very different effects of the sample dynamics on the reconstructed images depending on the FF-SRM method used. Notably, the MUSICAL algorithm appeared to exploit the signal fluctuations introduced via the sample dynamics, offering a greatly improved reconstruction of the mitochondrial outer membrane as compared to the stationary samples.

A considerable challenge for real samples, and especially for living samples, is the complete lack of ground truth. We can use what is known about the samples (e.g. the mitochondrial outer membrane is labelled) and our knowledge and experience with the different methods to aid our evaluation (e.g. circular mesh is a sign of failed MUSICAL reconstruction), the results will still be somewhat subjective and only useful until a certain point. If, for example, all the different methods showed different patterns of membrane domain proteins (only) in a plausible outer mitochondrial membrane area, we would have great difficulty in determining which one, if any of them, provided the correct picture of the membrane protein distribution. Therefore, simulations will be extremely important in the future development and evaluation of FF-SRM methods. They must, however, encompass sufficient complexity to be representative of real image data of dynamic and 3D biological systems [11]. This is not an easy task, but neither an impossible task in the current era of open science, global collaboration, and ever-expanding computational resources.

## Conclusion and outlook

We have seen that reliable reconstruction can be achieved for certain imaging conditions revealed via simulations of microscopy experiments. There are however still some challenges ahead for the young field of FF-SRM on the way towards reliable super-resolution image reconstructions from image sequences of densely fluctuating fluorophores for deriving useful biological inferences.

SOFI and MUSICAL were shown to have a different and superior ability to work with intensity fluctuations compared to other techniques. Both exhibited greatly improved reconstructions with longer image sequences and with higher rate of signal fluctuations. Lamentably, they also displayed the highest level of image degradation compared to the raw image data when the necessary requirements of the image data (like signal fluctuations) were not present. ESI, SRRF and SACD on the other hand, showed little to no improvement with the length of the image sequences and level of intensity fluctuation, but for all conditions gave less obvious artifacts and image structures that were usually more robustly in accordance with the conventional image data. As shown by simulations, the artifacts are still present and severe when considering details beyond the diffraction limit.

Choosing the right reconstruction parameters poses an additional challenge and introduces unwanted user subjectivity to the super-resolution images. The future development of these techniques should therefore encompass 'parameter-free reconstruction', intelligent pre-analysis of the data allowing for automatic selection of the optimal reconstruction parameters and feedback to the users about reconstruction quality and potential deficiencies of the image data. Examples of feedback include poor signal fluctuations, low SBR, sample is moving, sample appears out-of-focus, more frames needed, etc. The general lack of ground truth for living, dynamic samples is a substantial analytical challenge. Therefore, realistic 3D simulations of living cells (with known ground truth) will be important in the future development of these techniques.

We hope that this first comparative study of FF-SRM techniques highlighting the strengths and weaknesses of the different techniques will accelerate the arrival of a reliable and democratic nanoscopy technique suitable for a broad range of samples, likely combining strengths from the already suggested approaches. The potential rewards of true and reliable optical nanoscopy via conventional image sequences of 'any sample' together with the promising glints of reconstruction successes suggest that the many challenges along the way will be worth the effort.

## Data availability
The microscopy image data and reconstructions are available from https://www.3dnanoscopy.com/ff-srm-review/ or upon reasonable request. The password of the webpage is reviewpaper. This password protection will be removed when the manuscript is accepted and the data can be made public.

## Funding
IO and SA acknowledge strategic funding from UiT The Arctic University of Norway. KA acknowledges funding from a Horizon 2020 Marie Skłodowska-Curie Action (749666) and a Horizon 2020 ERC Starting Grant (804233). SA acknowledges funding from the Horizon 2020 Marie Skłodowska-Curie Action (749666). BSA acknowledges funding from RCN BioTek2021 (285571) and Marie Skłodowska-Curie grant agreement No. 766181. JC and NSB received funding from the European Union's Horizon 2020 research and innovation program under the Marie Skłodowska-Curie grant agreement No. 766181.

## Acknowledgements
The authors kindly thank Åsa B. Birgisdottir for providing cultures of fluorescently labelled H9c2 cells and pig heart tissue samples. Also, our gratitude to Mona Nystad for providing the placental sections and Randi Olsen for cutting all the ultrathin cryo-sections used in this study.

## Author contributions
KA conceived the idea. SA generated the simulation microscopy datasets and reconstructions for liposomes. JC produced and prepared the liposomes together with their size analysis. LV did the tissue sample preparation, acquisition and analysis. JC provided cultured macrophages for the fixed-cell experiments. ISO prepared, imaged and analyzed the live and fixed cell-culture samples. ISO wrote the draft manuscript and prepared the figures. KA, BSA and NSB supervised the project. Writing – review & editing, all authors. All authors have read and agreed to the submitted version of the manuscript.

## Ethics declarations

### Competing interests
The authors declare no competing interests.

### Ethical approval of tissue sections
Human placenta sections were collected at the University Hospital of North Norway according to the ethical protocol approved by the Regional Committee for Medical and Health Research Ethics of North

Norway (REK Nord reference no. 2010/2058-4). Written informed consent was obtained from the participant and the sample was treated anonymously. All methods were carried out in accordance with relevant guidelines and regulations.

Pig heart sections were obtained following the ethical protocols approved by the Animal Welfare Board at *UiT The Arctic University of Norway*, NO-9037 Tromsø, Norway, and the Norwegian Food Safety Authority *Mattilsynet*, NO-9008 Tromsø. All methods were carried out in accordance with relevant guidelines and regulations.

## Supplementary information
Supplementary results (pdf containing files) and the supplementary figure captions

Supplementary methods

# Supplementary methods for: *Fluorescence fluctuations-based super-resolution microscopy techniques: an experimental comparative study*

*Ida S. Opstad,[1, *] Sebastian Acuña,[1] Luís Enrique Villegas Hernandez,[1] Jennifer Cauzzo,[2] Nataša Škalko-Basnet[2], Balpreet S. Ahluwalia[1], Krishna Agarwal[1]*

[1]*Department of Physics and Technology, and* [2]*Department of Pharmacy, UiT The Arctic University of Norway, NO-9037 Tromsø, Norway.*

* ida.s.opstad@uit.no

## Methods

### Nanoscopy algorithms and their parameters

**ESI**, *Entropy-Based Super-Resolution Imaging,* estimates the likelihood of emitter molecule presence by calculating the local and cross-pixel (information-theory) entropy throughout the image sequence. The entropy can be thought of as the expectation value of the information content. The ImageJ-plugin used for analysis was retrieved from the software link available from the original publication [1].

Parameters: *number of images* in the output, number of *bins* for entropy, the *order* of the centralized moments. In addition, the plugin automatically uses maximum and minimum intensity in the image. The *number of images* can be useful when having long data sequences (thousands of frames) available for repeated ESI application on sub-stacks. The $0^{th}$ *order* is not applicable (vanishes), the $2^{nd}$ order is the variance, the $3^{rd}$ and $4^{th}$ *order* is related to the distribution shape and symmetry, and above $4^{th}$ order does 'not have an obvious correlation to the distribution itself' [1]. The *order n*, has a similar effect of raising the microscope PSF to the *n*th order, leading to a √n-fold narrowing of the approximated Gaussian signal. The default *order* value is 4, and that is what is used in this work unless otherwise stated. The *bins* were set to 100, and, unless otherwise stated, 1 image in output (as fairly short data sequences were used).

**bSOFI**, *balanced super-resolution optical fluctuation imaging* [2], is an improved SOFI implementation that achieves 'balanced' image contrast based on the emitters actually present, in contrast to the nonlinear response to brightness and blinking seen by the original SOFI auto- and cross-cumulant analysis [3]. The theoretical resolution improvement is √n, where n cumulant order. Mainly results using bSOFI has been addressed in this work. The analysis was performed in MATLAB using software copyright © 2012 Marcel Leutenegger et al, École Polytechnique Fédérale de Lausanne, under the GNU General Public License. No parameters are user defined.

**SRRF**, *super-resolution radial fluctuations* [4] is similar to SOFI but in addition assumes radiality of emitters. There is also a long range of additional options available in the SRRF plugin, that can be seen as different algorithms. The theoretical or expected resolution improvement depends on the reconstruction options, with the highest possible resolution improvement being by the temporal radiality auto-cumulant (TRAC) order 4 option, which is similar to SOFI order 4, so a possible √n = 2 - fold resolution improvement. Other possible options include the temporal radiality average (TRA) and temporal radiality pairwise product mean (TRPPM).

For the analysis presented in this work, multiple options were tried, primarily the 'default parameters' implying ring radius 0.5 with 6 axes, TRAC 2, with intensity weighing and 'minimize SRRF pattering'. When something different than default is indicated in the results, like TRAC 2, 3, or 4, it means the 'Advanced Settings' are activated with additional 'Gradient Smoothing' and 'Gradient weighting', together with the option indicated and the other default settings (like ring radius).

The images presented in the results, were the options yielding the best results out of the options tried. It should be noted, however, that it could be better options that were not tried out. Trying all possible combinations of parameter for SRRF analysis is beyond the scope of this work. The ImageJ-plugin used for the analysis was retrieved from link in the original publication [4].

**SACD**, *super-resolution method based on autocorrelation two-step deconvolution [5],* is (also) inspired by SOFI, but in addition to signal autocorrelation performs a (Lucy-Richardson) deconvolution step before and after the autocorrelation analysis. The autocorrelation analysis is based on an algorithm they call *multi-plane auto-correlation* (MPAC), to counter a typical lack of image frames for autocorrelation analysis by also correlating combination of frames in multiple steps (or planes). It is especially developed for live-cell imaging under the conditions of low signal and few input frames (e.g. they use 16 frames in their article).

The reconstructions made using SACD was performed in MATLAB environment using the source code following link in the original publication. The tiff files were converted to mat-format in MATLAB prior to the SACD computation. After reconstruction, the data was converted to tiff files using the MATLAB *save* function.

Parameters: *number of deconvolution iteration steps.* 10 steps were used for the results presented. 20 steps were tried for some computations, but no significant improvement was found. *Magnification:* image upscaling by Fourier interpolation. 8 were used for the results presented. *Power of PSF* for the second deconvolution step: PSF powers were tried in the range 2-6. *Orders of MPAC used:* 2,3,4,5 and 6. The PSF order was the same as for the MPAC for the presented results. The order used for particular results is indicated in the figures. Optical parameters for PSF calculation: *wavelength, numerical aperture and pixel size.* These were in accordance with the particular sample dataset being processed.

**MUSICAL**, *Multiple Signal Classification Algorithm [6]*, calculates higher resolution through singular value decomposition and the resulting eigenimages. Each eigenimage represents a particular pattern found in the image stack, and the associated singular value describes how significant or likely the pattern is. The larger the singular value of an eigenimage, the more likely this is to be an underlying image feature rather than resulting from noise or background signal. This way, by selecting a *threshold* for what are signal and what are noise (or unlikely) image features, the MUSICAL image is computed by taking the ratio of an independent estimate for both the signal and the noise. This ratio is for the final image intensity value taken to the power of the parameter *alpha* (usually 4). The computations are done on sub-images whose size depends on the imaging system's PSF taken into account.

The MUSICAL image and singular value computations were done in ImageJ using a plugin and macro for multicolor time-lapse processing. Both the plugin and macro are available from https://github.com /sebsacuna/MusiJ. The required input <u>parameters</u> are *pixel size*, *subpixels* (image magnification factor), (microscope system) *magnification*, *numerical aperture*, *alpha*, and *batch size* (number of images used for the MUSICAL image computation). For each color channel one must also specify *emission wavelength* and *threshold*. The threshold is a parameter particular to MUSICAL and determines the cutoff for what fluctuation eigenvectors goes into signal or noise space. This parameter was selected experimentally by trying values in the range of the second singular values (middle, lower and high end of the spectrum).

Unless stated otherwise in the results, the following parameters were used: emission wavelength 510 nm, pixel size 80 nm, subpixels 10, magnification 1 (already calculated for in the pixel size), alpha 4, numerical aperture, 1.420. The batch size was in the range 50 to 400 (stated for individual results).

Different threshold values were used (stated in the Results section), but were in general chosen from the range of the second singular values (as described in [7]), where the middle of this range was usually found the best.

**HAWK,** *Haar wavelet kernel* [8] is not a super-resolution method in itself, but a data pre-processing technique that can be used to increase data sparsity by distributing the data over more frames. The motivation behind this technique is that high-density data is difficult to analyse and leads to mislocalizations and image artefacts using e.g. traditional super-resolution techniques like single molecule localization microscopy. In the manuscript, they show improved image reconstruction also for SOFI and SRRF, relevant to the current work.

Different options available for this ImageJ plugin (retrieved from link in [8]) are: Settings: *number of levels* (3,4 or 5), negative values: *separate* or *absolute value*, Output order: *group by level* or *group by time*. Unless stated otherwise in the results, the following parameters were used: Settings: number of levels: 5, negative values: separate, Output order: group by level. These parameters were chosen from initial experimentation and based on results in the HAWK publication.

HAWK pre-processing is presented for ESI, SRRF, bSOFI and for MUSICAL, but not for SACD because of MPAC large memory requirements for the extended datasets and poor performance under initial testing (degradation of results compared to SACD alone).

## Data analysis

It can be difficult to objectively evaluate the reconstructed images. Resolution is not merely the width of slim line profiles or visible separation between two closely spaced bright spots in the image. These structures must also represent the actual underlying sample, and not noise or reconstruction artifacts. For the simulated data, evaluation can be done fairly objectively as the ground truth is known. For the live-cell data on the other hand, the ground truth is not available. The evaluation was done via visual inspection following these criteria (-/+ indicate bad or good):

(-) In image areas where clearly nothing but noise or background was present, reconstructed structures are artifacts.

(-) Occurrence of suspect pattern of unlikely biological origin. Especially if these shapes change for different reconstruction parameters.

(+) Images reveals null or close to none signal in 'no object areas', but significant signal in 'object of interest' area.

(+) Image structures are consistent for different parameters (if available).

(+) Images reveal sub-resolution limit structures on object of interest areas and nowhere else. It is good if these structures are in accordance with what is already known about these structures/organelles, although this cannot be strictly required as most nanoscale cellular structures in living cells cannot be strictly assumed to have the same nanostructure as seen by electron microscopy (EM) of fixed, starkly treated cells.

(+) Excludes out-of-focus structures, rather than producing artefacty, nonsense sample details.

Other relevant aspects are ease of use and reconstruction time.

The signal to background ratio (SBR) was measured from the mean of small regions where samples were present or not. This is a difficult measure to make, as it varies throughout the image, and it is not always obvious from the images where the object of interest is or not (especially in the case of the dense mesh of the ER). Measuring the background where no objects are visible is neither fair, as for example objects in a different imaging z-plane will contribute strongly to the background and heavily challenge the reconstruction of the in-focus objects.

The intensity line profiles were measured in Fiji/ImageJ using a line width of 1. All other image processing tasks (like gamma intensity adjustment, image summation or standard deviation, insertion of scale bar etc.) were also performed in Fiji/ImageJ (version 1.52p).

## Data acquisition and sample preparation

### Imaging system

The data was acquired using a commercial OMX V4 optical microscope with 3 cameras and up to four-channel imaging. The objective lens was a 60X 1.42NA oil immersion lens, except for the TIRFM data, where the objective was 60X 1.49NA TIRF lens. Widefield, epifluorescence single-plane time-lapse data was acquired in sequential imaging mode of the different color channels to avoid blead-through between the channels.

### Liposomes

Liposomes were prepared according to the film hydration method [9]. Soy phosphatidylcholine (SPC; generously provided by Lipoid GmbH, Ludwigshafen, Germany) was used as main lipid ingredient in concentration of 10 mg/mL. 1-myristoyl-2-[7]-sn-glycero-3-phosphocholine (N; Avanti Polar Lipids, AL, USA) was chosen as fluorescent marker in 0.03 mg/mL. Excitation and emission wavelength of the incorporated fluorophore are 476 nm and 537 nm respectively.

Both lipid ingredients were dissolved in methanol and dried to a thin film through low-pressure rotary evaporation (Büchi Rotavapor R-124, Büchi Labortechnik, Flawil, Switzerland). Distilled water was used to hydrate the lipid film and form large multi-lamellar vesicles. After overnight stabilization, the vesicle size was reduced combining sonication and sequential hand extrusion through polycarbonate membranes of 400, 200 (and 100) nm sieving sizes (Whatman Nucleopore™) to the target size of A) 250 nm and B) 100 nm. The size distribution was derived from a Gaussian-like fitting of dynamic light scattering signal (Malvern Zetasizer Nano – ZS, Malvern, Oxford, UK) and resulted in A) 240±77 nm and B) 117±30 nm. The respective polydispersity indexes (PdI) of 0.410 and 0.217 described A as a polydispersed system (PdI>0.25) and B as a monodispersed one (PdI<0.25). ζ-potential measurements obtained through laser doppler electrophoresis (Malvern Zetasizer Nano – ZS, Malvern, Oxford, UK) are included for completion, Table S1.

Table S1: Standard processing and characterization of liposomal formulation.

| N-Lip FORMULATIONS | | | | | | |
|---|---|---|---|---|---|---|
| | Composition | | Processing | | Characterization | |
| N-Lip | SPC conc | N conc | Sonication | Extrusion | Size | | ζ-potential |
| | [mg/mL] | [mg/mL] | sec | [passes]x[nm] | [nm] | PdI | [mV] |
| A | 10 | 0.03 | 120 | 4x400 + 4x200 | 240±80 | 0.410 | -8.0±3.7 |
| B | 10 | 0.03 | 120 | 4x400 + 4x200 + 4x100 | 117±30 | 0.217 | -6.4±2.7 |

Both liposomal suspensions were diluted 1:1000 in distilled water to the final lipid concentration of 10 µg/mL (and fluorophore concentration to 0.03 µg/mL). 3 µL droplets were then placed on ethanol-cleaned coverslips and covered with thin (2 mm) patches of solid agarose (2 % in water) for immobilization.

## Fixed-cells

Murine macrophages RAW 264.7 (ATCC® TIB-71TM, ATCC, Manassas, USA) were cultured in RPMI-1640 medium, supplemented with 1% penicillin-streptomycin and 10% fetal bovine serum. The cell culture in a 25-cm2 flask was incubated at 37 °C and 5% $CO_2$ until 100% confluence. Cells were then scraped, counted in a Neubauer chamber and diluted to 50,000 cells/mL, prior to the final plating into 35-mm petri dishes – No. 1.5 coverglass (MatTek Corporation, Ashland, USA).

The cells were fixed at a confluence of about 50% at room temperature (RT) using 4% PFA in Cytoskeletal buffer [10]. The samples were subsequently washed in PBS and labelled with Phalloidin-ATTO647N (Merck) using 3µL/100µL PBS for 1h at RT, and CellMask™ Orange Plasma Membrane Stain (ThermoFisher Scientific) using a concentration of 1 to 2000 in PBS for 5min. The samples were washed repeatedly in PBS and finally immersed in PBS for imaging.

## Tissues

### Collection and preservation

The samples were collected and preserved following the Tokuyasu method for cryo-sections [11]. Human chorionic tissue from full-term placenta was dissected immediately after delivery into 1 $mm^3$ blocks, rinsed in 9 mg/mL NaCl and transferred to 1X PHEM-buffer. The blocks were incubated in 8% formaldehyde in PHEM-buffer at 4 °C overnight and immersed for 1 h in 0.12% glycine at 37 °C. Thereafter, the samples were infiltrated with 2.3M sucrose at 4 °C overnight, and mounted on specimen pins before storage in liquid nitrogen. Analogously, myocardial samples were collected from anesthetized pigs using a biopsy needle. The cardiac samples were dissected, washed and further prepared in identical manner as with the placental sample.

### Fluorescent staining

The samples were prepared for microscopy as in [12]. In summary, the tissues were cryo-sectioned with an EMUC6 ultramicrotome (Leica Microsystems, Vienna, Austria), collected with a wire loop filled with a 1:1 pick-up solution of 2.3 M sucrose and 2% methylcellulose, and placed onto poly-L-lysine coated #1.5 high-precision coverslips. Subsequently, the cryo-sections were washed 3 × 7 min with PBS at 4°C and fluorescently labelled according to the experimental plan at RT. The placental tissue was incubated for 15 min in a 1:100 solution of phalloidin-ATTO647N in PBS, washed 2 × 5 min with PBS, incubated in a 1:2000 mixture of CellMask Orange in PBS for 10 min, and washed 2 × 5 min with PBS before mounting. Analogously, the cardiac tissue was labelled with CellMask Orange for 10 min, followed by 2 washing steps of 5 min in PBS. Thereafter, the labelled sections were mounted onto standard microscope glass slides (placenta) or a reflective silicon chip (heart) using Prolong Gold and sealed with nail varnish. The samples were stored at 4°C and protected from the light before imaging.

## Live-cell data

Rat cardiomyoblast cells (H9c2) were cultivated in glass bottom petri dishes (MatTek Corporation, Ashland, USA) and transiently transfected with organelle (mitochondria and endoplasmic reticulum KDEL) targeted fluorescent fusion proteins (OMP25-mCherry and KDEL-EGFP). Imaging was done in cell-culture medium (DMEM with 10% fetal bovine serum), using heating and an environmental chamber set at 37°C and 5% $CO_2$.

## Simulated data

The simulations of samples of actin and mitochondria (tori) were generated using the same optical parameters: 510 nm emission wavelength, 1.42NA, 80 nm pixel size of the projected optical image. The photokinetic model for the fluorophores considers two states: emitting and non-emitting, with the time spent in each state modelled as a random variable with Poisson distribution of mean $\tau$. While emitting, the emission rate is constant for each emitter and the intensity at every frame is computed accordingly. The Gibson-Lanni model [13] was used as model for the point spread function to generate the noise-free microscopy image sequences. To these images, the noise was added by normalizing the noise-free image to the range [0,1] and then using a linear transformation with parameters $a = 200$ and $b = 50$ using the formula below. The parameter $a$ corresponds to the maximum signal, while $b$ is an offset used to replicate the camera behavior. The resulting intensities are the mean of a Poisson distribution. Therefore, the final image intensity for each pixel is generated from the equation $i_f = Poisson(ai + b)$. The slope $a$ is related to the maximum signal value, while the intercept $b$ is the offset of the camera.

The actin sample is formed by single tubes of diameter 6 nm, with emitters randomly placed on the surface with a linear density of 500 emitters per µm. A linear density is used since the size of the strand is an order of magnitude smaller than the pixel size. In the mitochondria case, the sample is constructed tori distributed as a grid of 2 rows and 3 columns. Each row contains tori of the same size while each column represents a single axial position referred to as z, which is the distance between the center of gravity of the tori and the coverslip. Each torus is built as solid of revolution, by rotating a small circle around a larger one. The upper row contains tori with minor radius of 100 nm and major radius 300 nm, while the structure in the row at the bottom is 200 nm and 500 nm respectively. Three planes with axial positions from the coverslip 500 nm, 700 nm and 900 nm (going from left to right in the sample), 700 nm being the plane in focus. The emitters are located on the surface of the tori with a density of 400 emitters per µm².

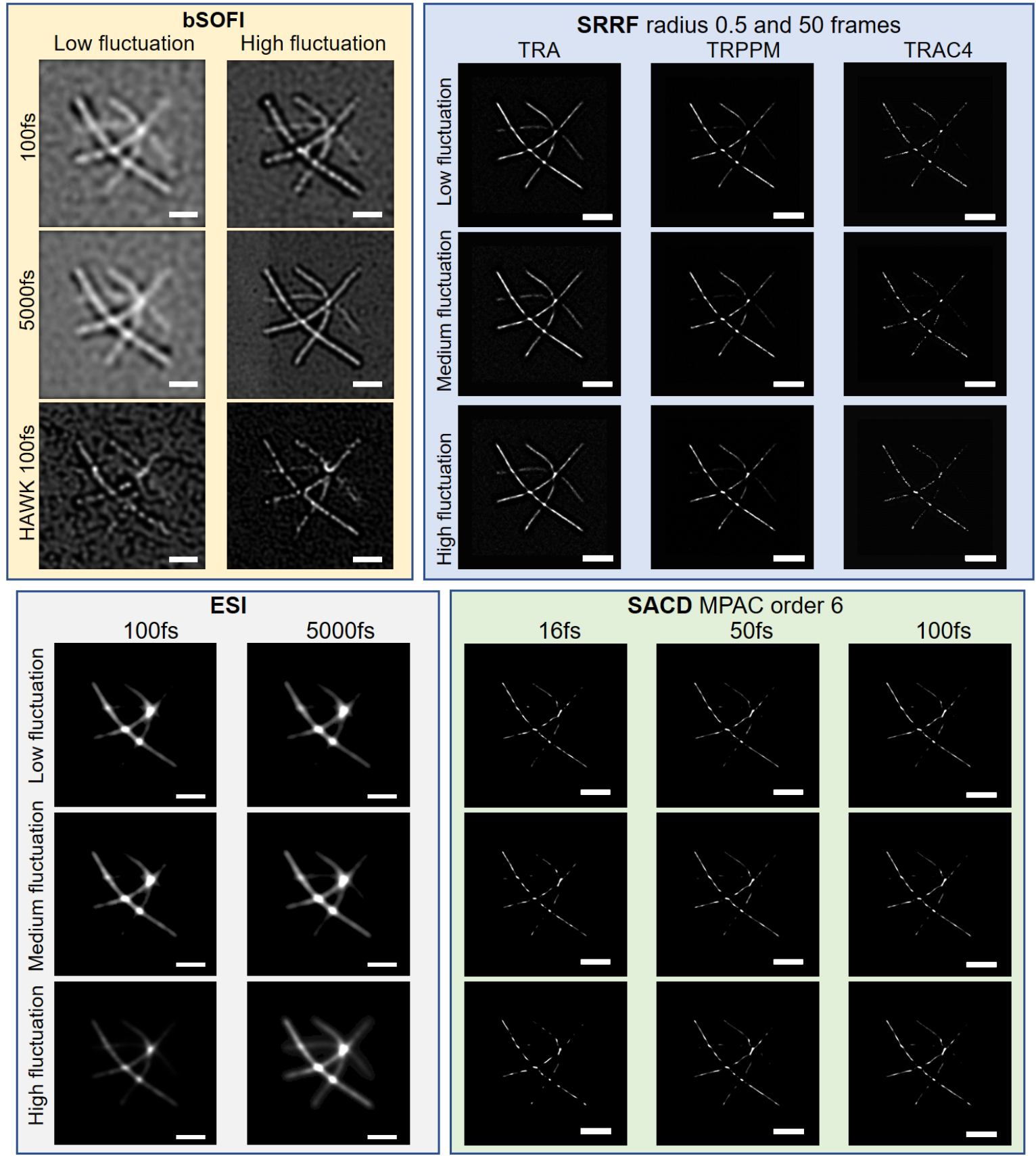

**Figure S1**

*Simulated 3D actin strand:* **bSOFI** *for two different fluctuation levels, HAWK and different number of frames (fs). The bSOFI images improve drastically by using both HAWK and a higher number of frames, but only for the higher fluctuation level. The scale bars are 1 µm;* **SRRF** *for different reconstruction options and fluctuation levels. The temporal radiality average (TRA) includes most of the out of focus strands and also gives higher background signal for higher fluctuation level. The temporal radiality pairwise product mean (TRPPM) and temporal radiality auto-cumulant order 4 (TRAC 4) option, appears for this data similar and largely unaffected by the fluctuation level for this data. The actin strand reconstructions from TRAC 4 are slimmer than for the other options, which can be explained by TRAC 4 being the SRRF options with the highest theoretical resolution improvement, namely a factor of 2 (Norder = 2 in the same manner as for the SOFI orders);* **ESI**: *the results for different number of frames and different fluctuation levels are very similar for ESI, except more out-of-focus structures being visible for higher fluctuation levels. To better visualize the finer image details alongside the bright spots, the ESI images are γ=0.5 intensity adjusted;* **SACD**: *The images for different number of frames and fluctuation level are for SACD nearly indistinguishable, but for the highest number of frames and highest fluctuation level, the inclusion of out-of-focus strands is slightly higher than for the remaining images. The scale bars are 1 µm.*

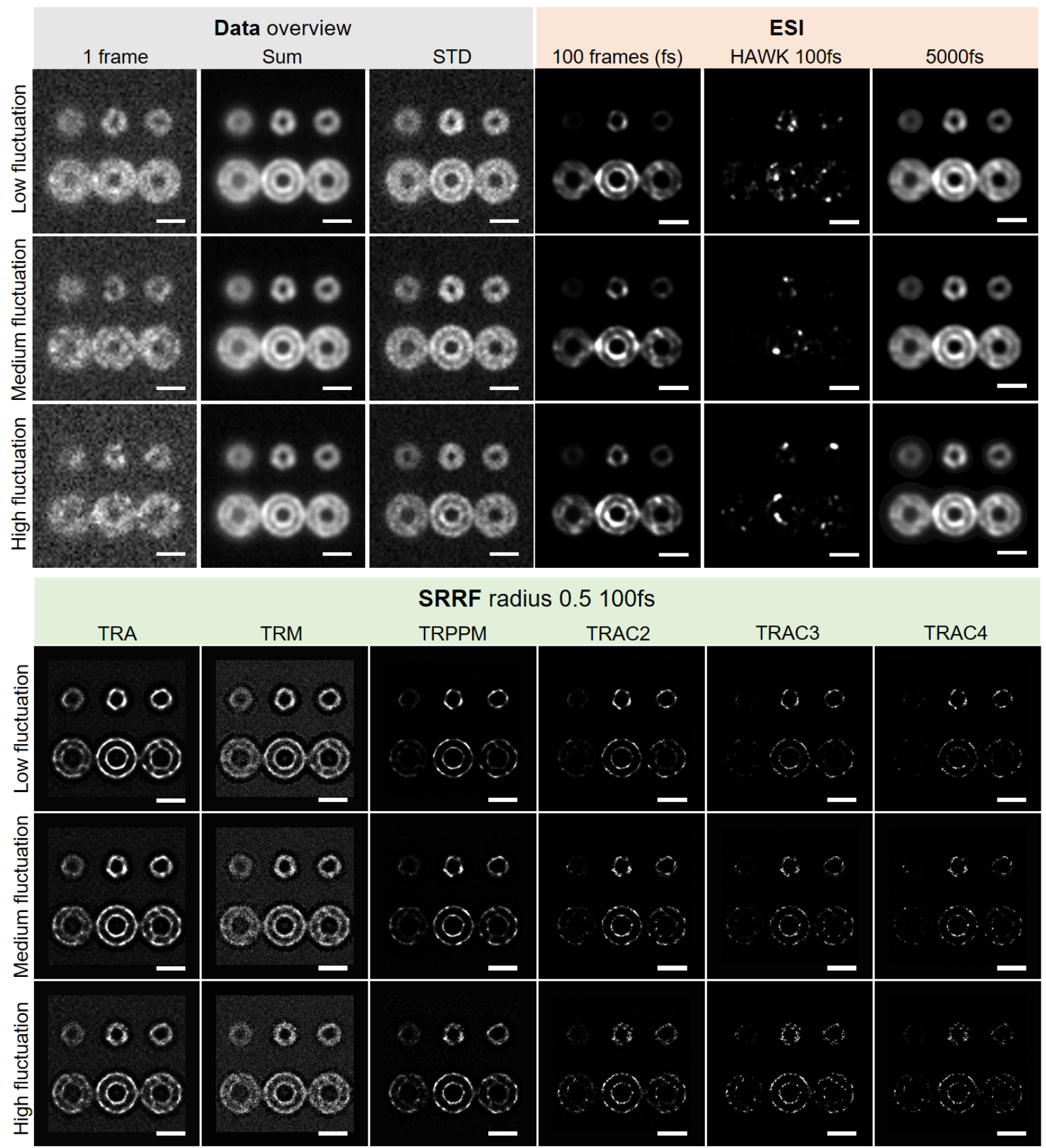

**Figure S2**

*Simulated 3D tori (doughnuts/mitochondria):* **Data** overview of the three different fluctuation data sets, each containing a single frame, a sum image of 100 frames, and the standard deviation image of 100 frames (STD); **ESI**: effect of fluctuation level, HAWK and number of frames. While HAWK significantly degrades the ESI image, changing the fluctuation level or the number of frames does not significantly alter the ESI results; **SRRF**: effect of using different options for SRRF reconstruction. The effect on the SRRF images from of changing reconstruction parameters (or *options*) is much larger than the effect of fluctuation level. Although the double ring from the larger torus is visible for any fluctuation level, the width of the doughnuts is more accurately represented with higher fluctuation level. The inability to resolve the inner and outer rings of the smaller doughnuts might be at least partly due to the poorer z-sectioning than MUSICAL or ESI (See Figure 2 in the main manuscript). The scale bars are 1 μm.

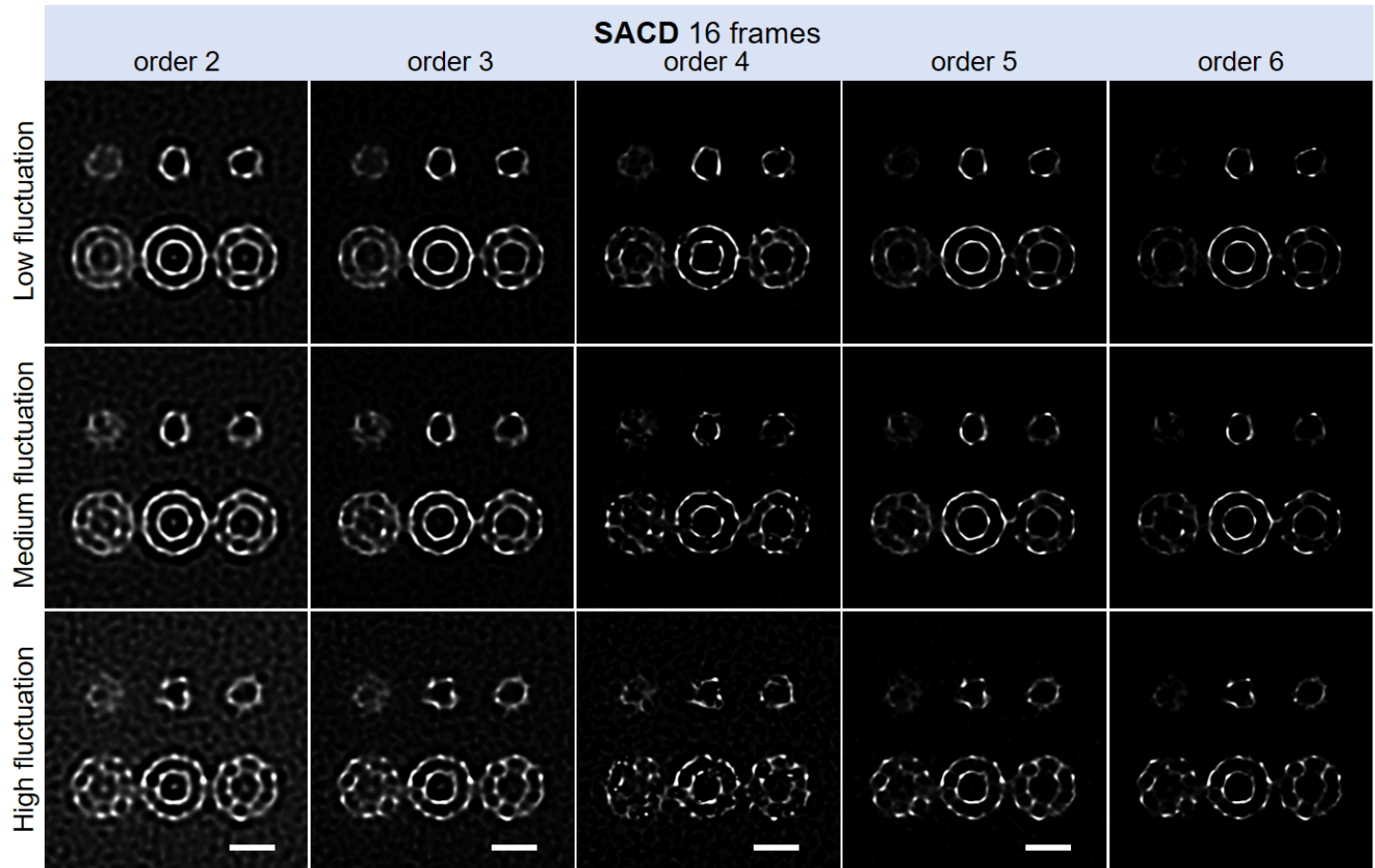

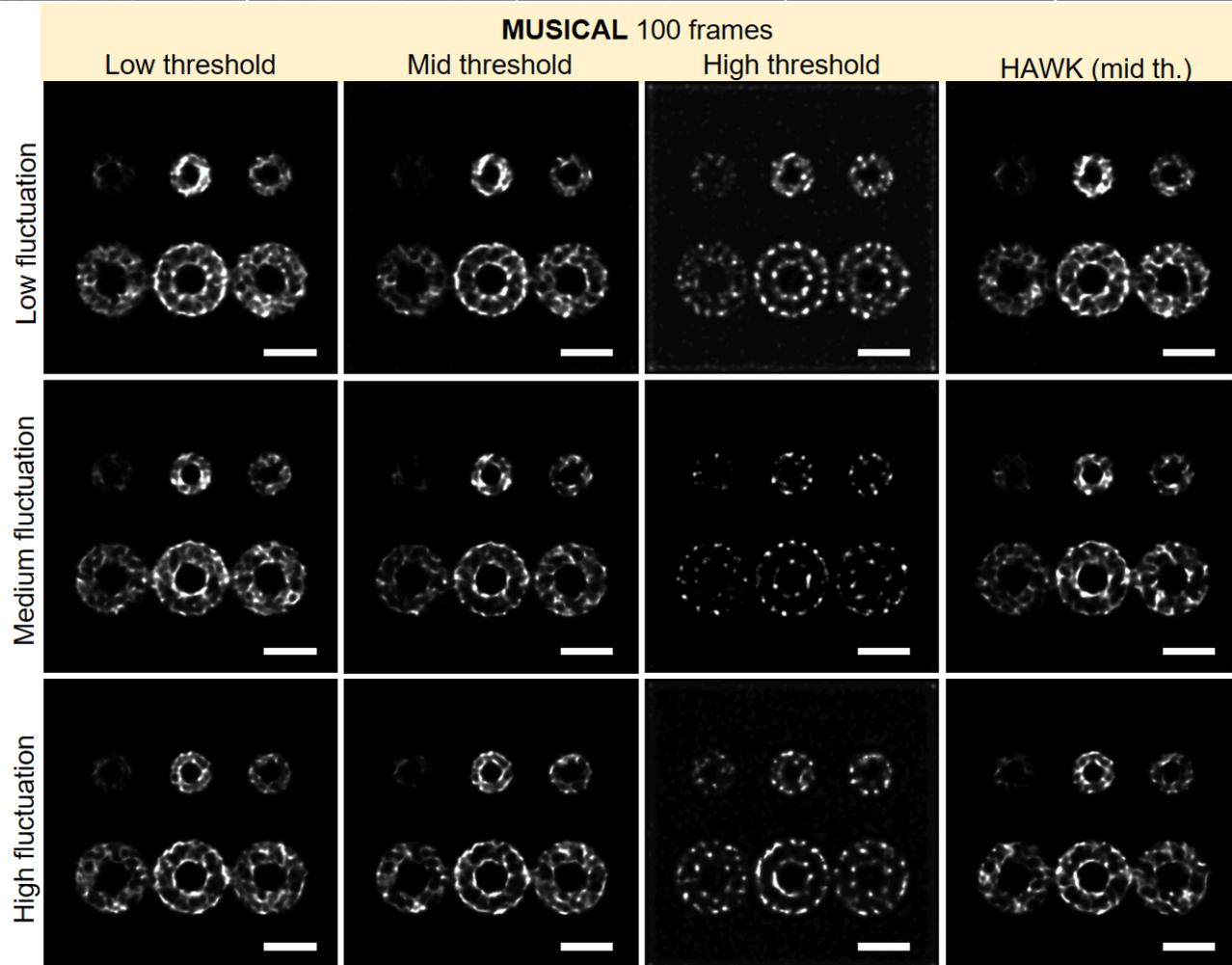

**Figure S3**

*Simulated 3D tori:* **SACD**: effect of fluctuation level and the MPAC (multiplane autocorrelation) order. The SACD images become worse and appear more affected by noise for higher fluctuation levels. The higher the MPAC order, the better the noise rejection and the higher rejection of out of focus structures (possibly also in-focus structures, see Suppl. Figure S 13 EPI SACD); **MUSICAL**: effect of different fluctuation levels and threshold parameters. The resolving power of MUSICAL becomes clearly better with higher fluctuation levels. Low, mid and high threshold are referring to the range of 2nd singular values according to the MusiJ plugin [1]. The lower the threshold, the higher portion of the signal fluctuations are included as 'signal' and less as 'noise'. This increases the information available for calculating a nanoscopy image, but also increases the contribution of potential noise and out-of-focus signal which can lead to image artifacts. HAWK preprocessing lead to similar results for MUSICAL as without HAWK.

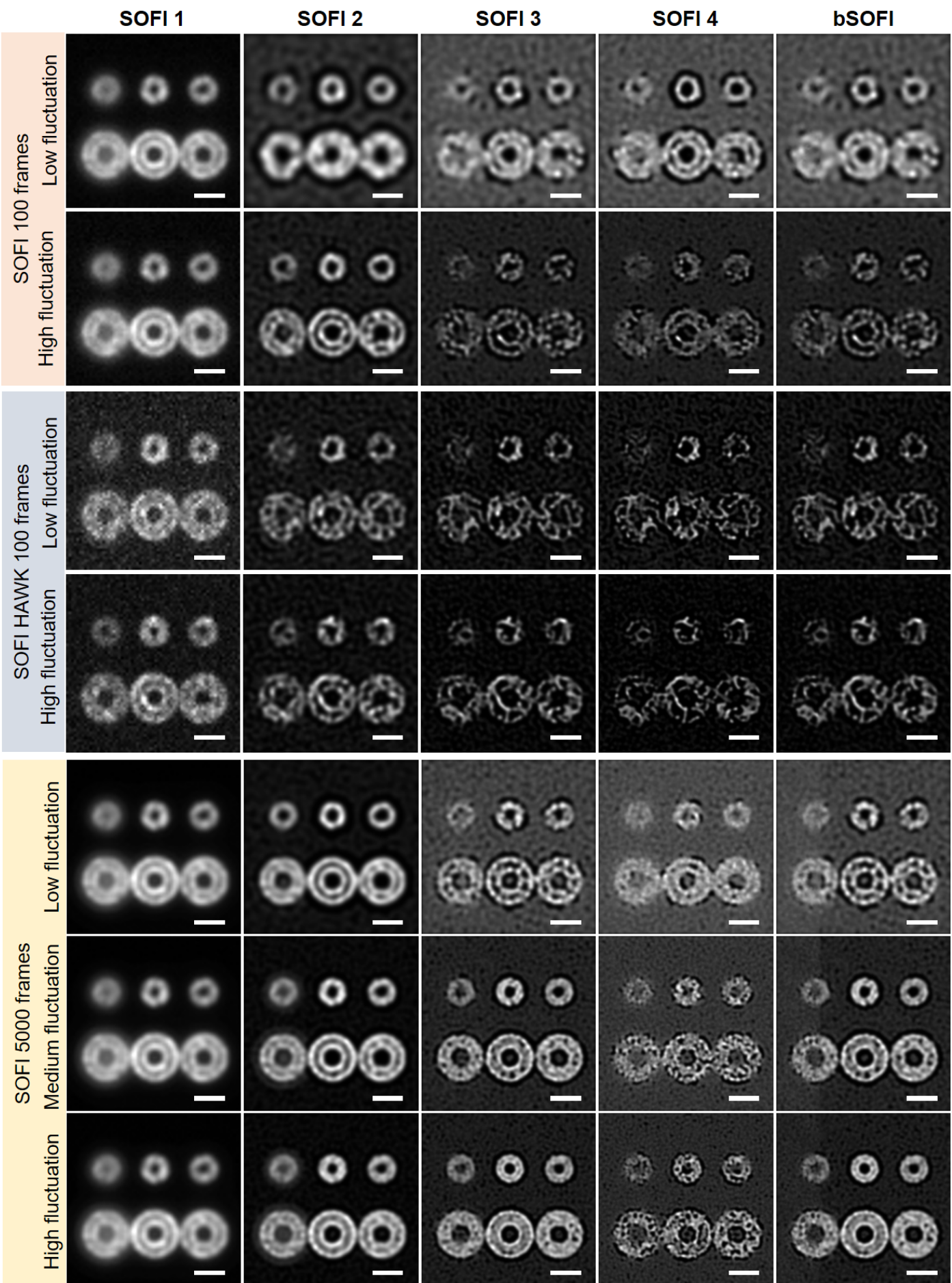

**Figure S4**

*Simulated 3D tori:* effect of fluctuation level, HAWK, and number of frames for different SOFI orders and bSOFI. The first two SOFI order, SOFI 1 and SOFI 2, are equivalent to the mean and variance images, respectively, and do not offer theoretical resolution improvement beyond the diffraction limit. SOFI 3 (third order), on the other hand, does possess theoretical super-resolving capabilities (up to a factor of $\sqrt{3} \approx 1.73$), something which is experimentally confirmed from the in-focus tori (middle column) from the highest fluctuation level and frame number (bottom row), where the double ring characteristics of also the smallest torus can be clearly visualized. SOFI 4 appears here to capture primarily sample noise rather than further resolution increase. The bSOFI images are for these data nearly identical to SOFI 3.

# Assessment of the number of frames parameter for 100 nm liposomes

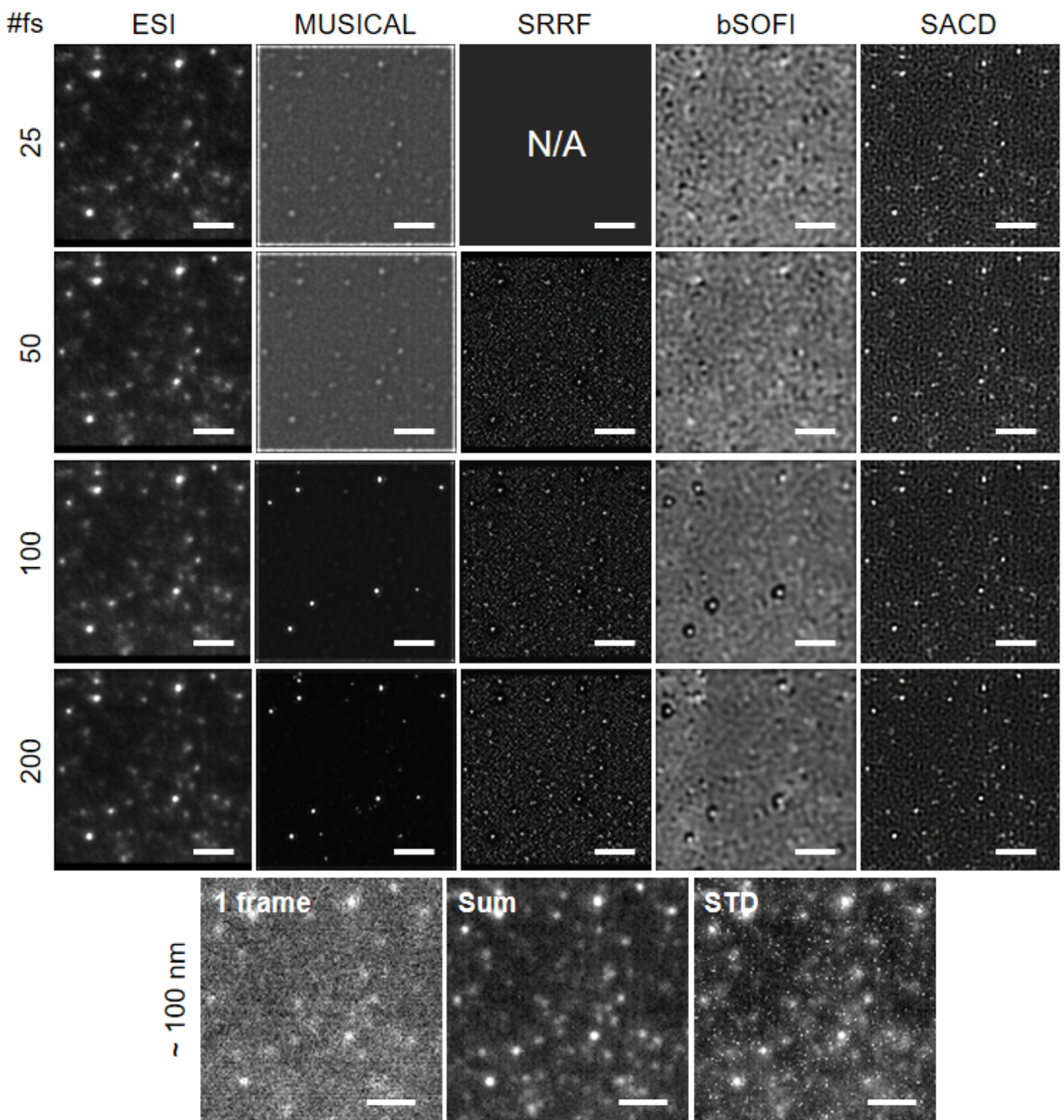

**Figure S5**

*Liposomes 100 nm:* assessment of the number of frames (#fs) parameter and an overview of the data, containing a single frame, a sum, and the standard deviation image (STD). The scale bars are 2 μm. 100 frames were chosen for further analysis, exhibiting reduced background and high visibility of image features appearing to be in-focus liposomes in the MUSCAL and bSOFI images, although the remaining techniques appear similar for all frame numbers.

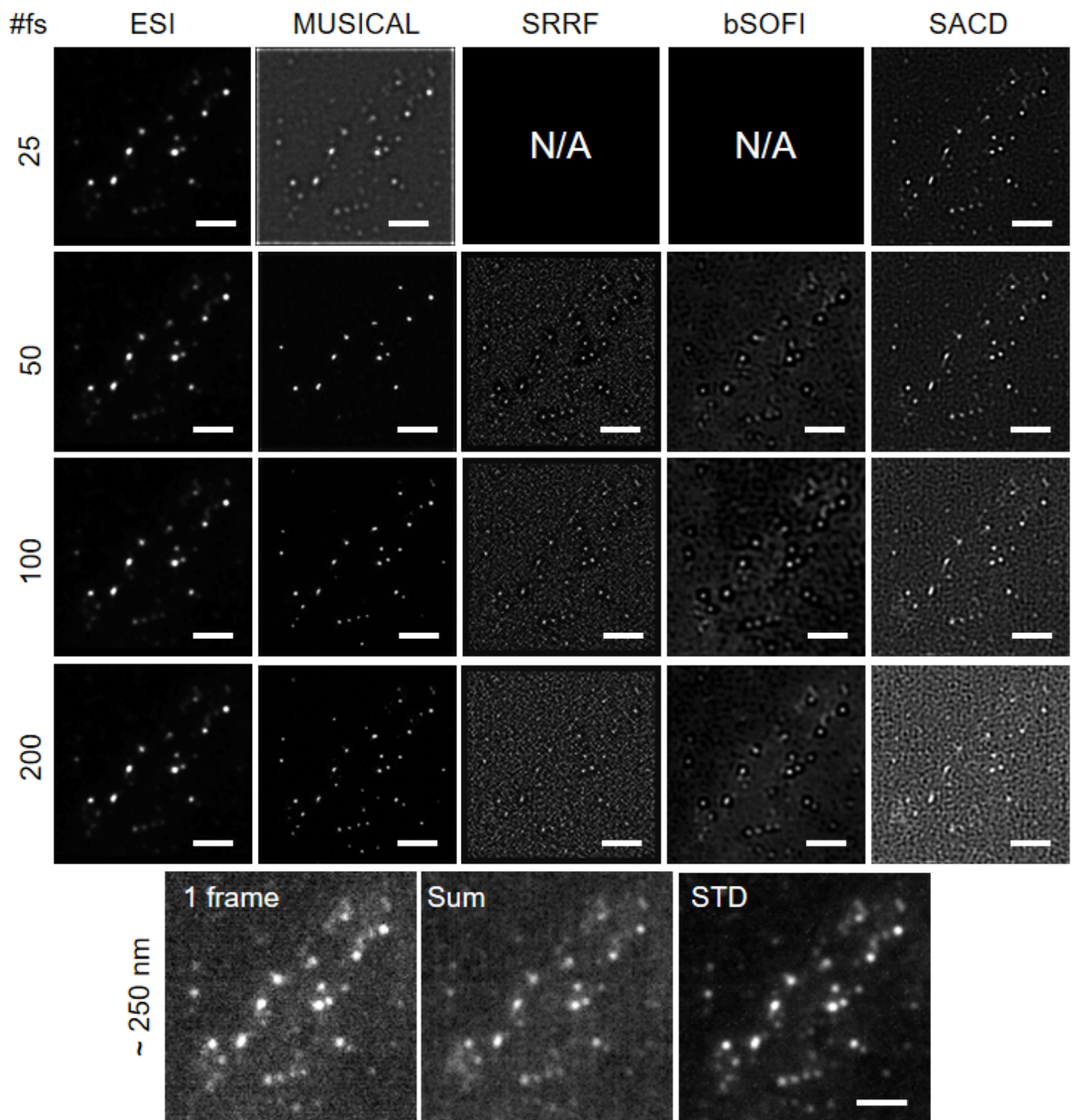

**Figure S6**

*Liposomes 250 nm:* assessment of the number of frames (#fs) parameter and an overview of the data, containing a single frame, a sum, and the standard deviation image (STD). The scale bars are 2 µm. The number of frames selected in most cases (and 'default' in case of doubt) was 100. A different frame number was chosen only in the of SACD (25 fs) and bSOFI (200 fs) for the 250 nm sample, as these results were evaluated significantly better, yielding less background artefacts.

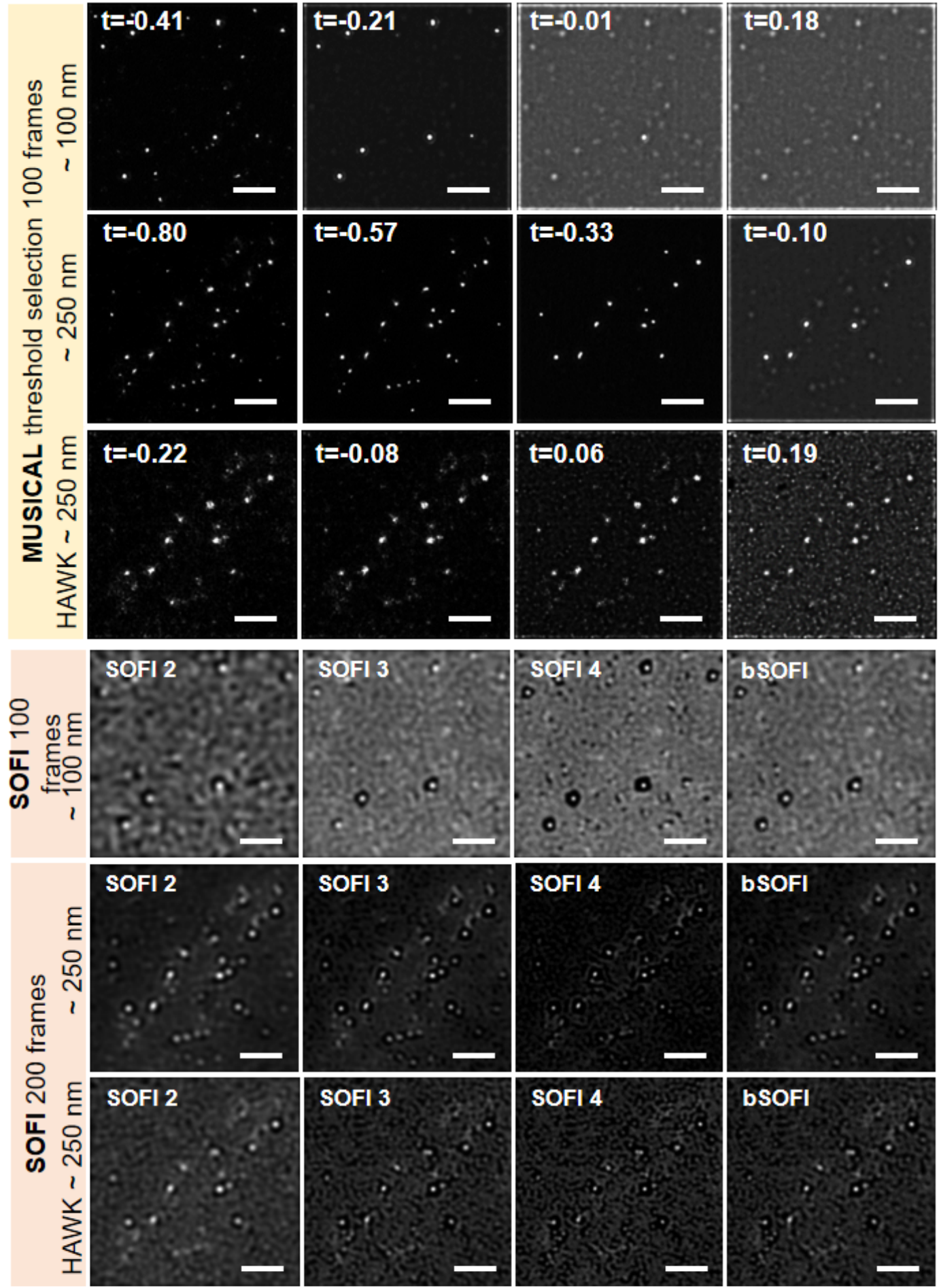

**Figure S7**
*Liposomes 100 nm and 250 nm:* assessment of the threshold parameter for MUSICAL and best SOFI method for the assessed 100 nm and 250 nm liposome data. For MUSICAL, the thresholds t=-0.21 (100 nm) and -0.57 (250 nm) were found best and chosen for further analysis. For SOFI, bSOFI was found best. HAWK increased the effect of the background caused by the agarose autofluorescence signal. The scale bars are 2 µm.

| 100 nm | SUM [nm] | ESI [nm] | MUSICAL [nm] | SRRF [nm] | SOFI [nm] | SACD [nm] |
|---|---|---|---|---|---|---|
| 1 | 366 | 207 | 33 | 54 | 183 | 139 |
| 2 | 324 | 188 | 49 | 45 | 175 | 136 |
| 3 | 328 | 178 | 43 | 88 | 231 | 153 |
| 4 | 314 | 218 | 42 | 49 | 230 | 165 |
| 5 | 285 | 188 | 36 | 53 | 188 | 131 |
| MEAN | 323 | 196 | 41 | 58 | 201 | 145 |
| STD | 29 | 16 | 7 | 17 | 27 | 14 |

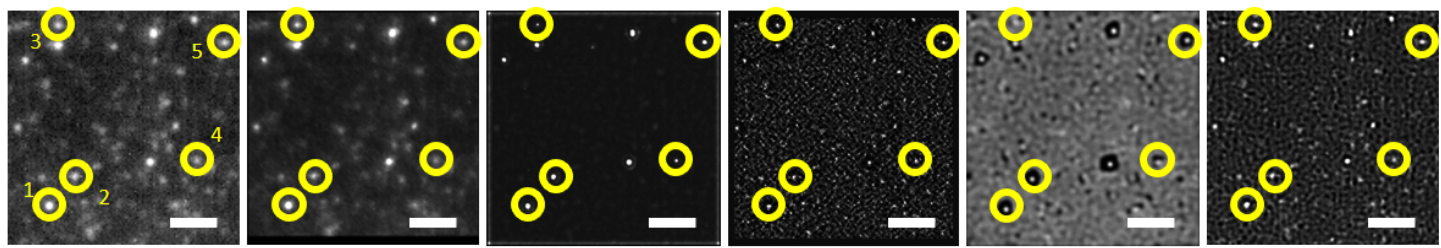

| 250 nm | SUM [nm] | ESI [nm] | MUSICAL [nm] | SRRF [nm] | bSOFI [nm] | SACD [nm] |
|---|---|---|---|---|---|---|
| 1 | 290 | 166 | 44 | 50 | 156 | 98 |
| 2 | 309 | 195 | 47 | 44 | 202 | 135 |
| 3 | 454 | 206 | 58 | 49 | 198 | 132 |
| 4 | 269 | 198 | 68 | 44 | 189 | 107 |
| 5 | 304 | 150 | 51 | 47 | 174 | 85 |
| MEAN | 325 | 183 | 54 | 47 | 184 | 111 |
| STD | 74 | 24 | 10 | 3 | 19 | 22 |

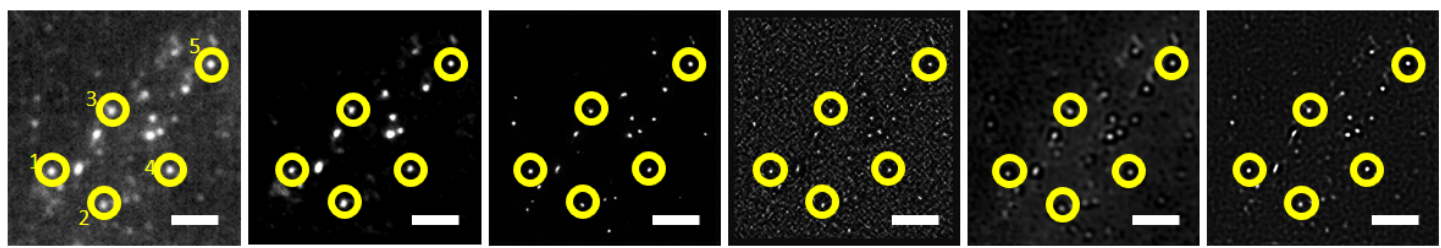

**Figure S8**

*Table of size measurements for the 100 nm and 250 nm liposomes.* The measured particles are indicated in the panels below the respective tables. The scale bars are 2 µm. The number of frames were 100, except for the cases of SACD (25 fs) and bSOFI (200 fs) for 250 nm liposome. A different number of frames were used in these cases based on the number of frames study (Suppl. Figure S6).

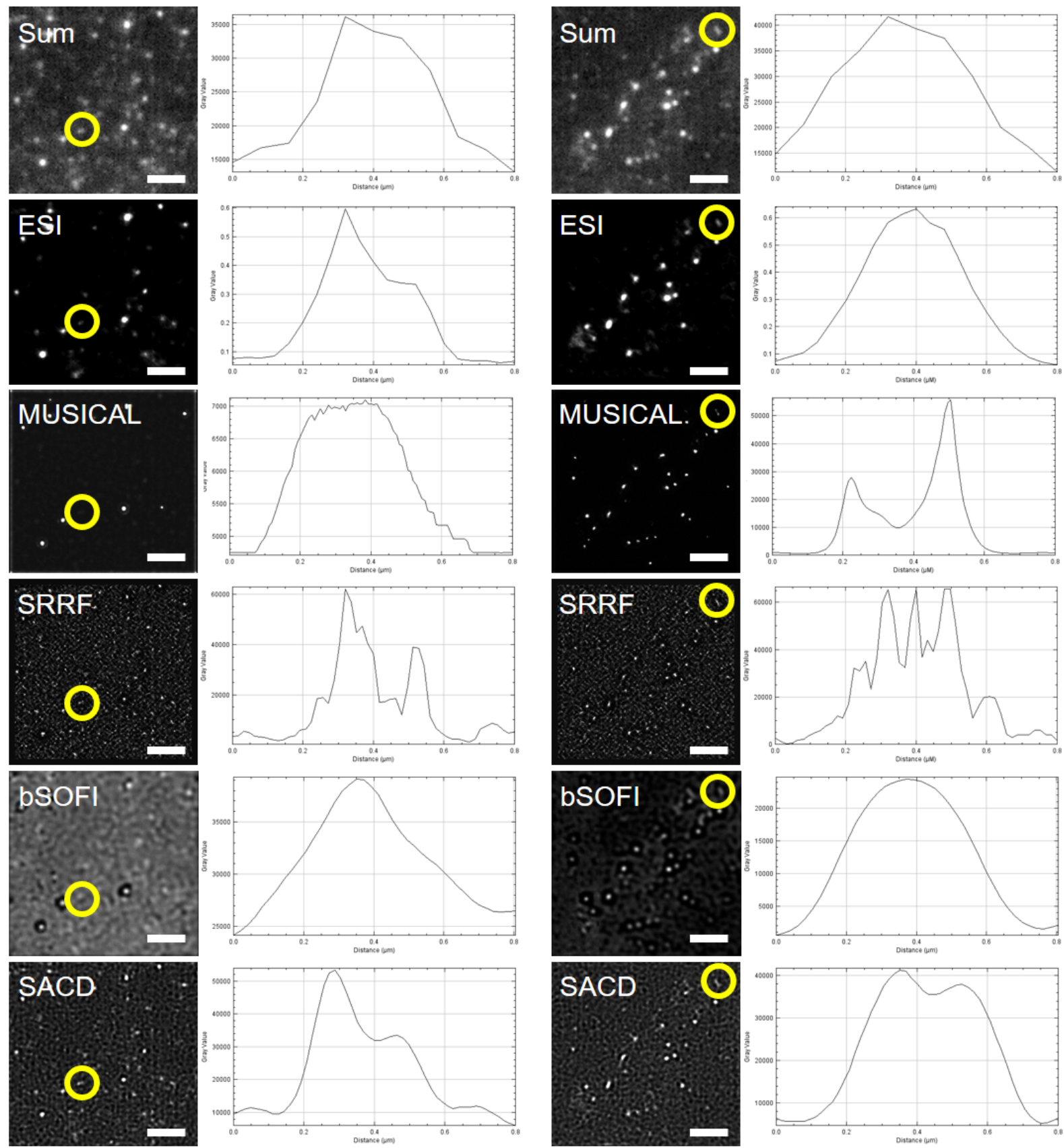

**Figure S9**

*Resolution measurements for the 100 nm and 250 nm liposomes via line profiles.* The regions containing elongated spots (indicating presence of multiple liposomes) are indicated by the yellow circles. The double dip seen by some plots can be a sign resolution enhancement. However, high prevalence of reconstruction artifacts (partly due to agarose autofluorescence), renders the measurements unreliable.

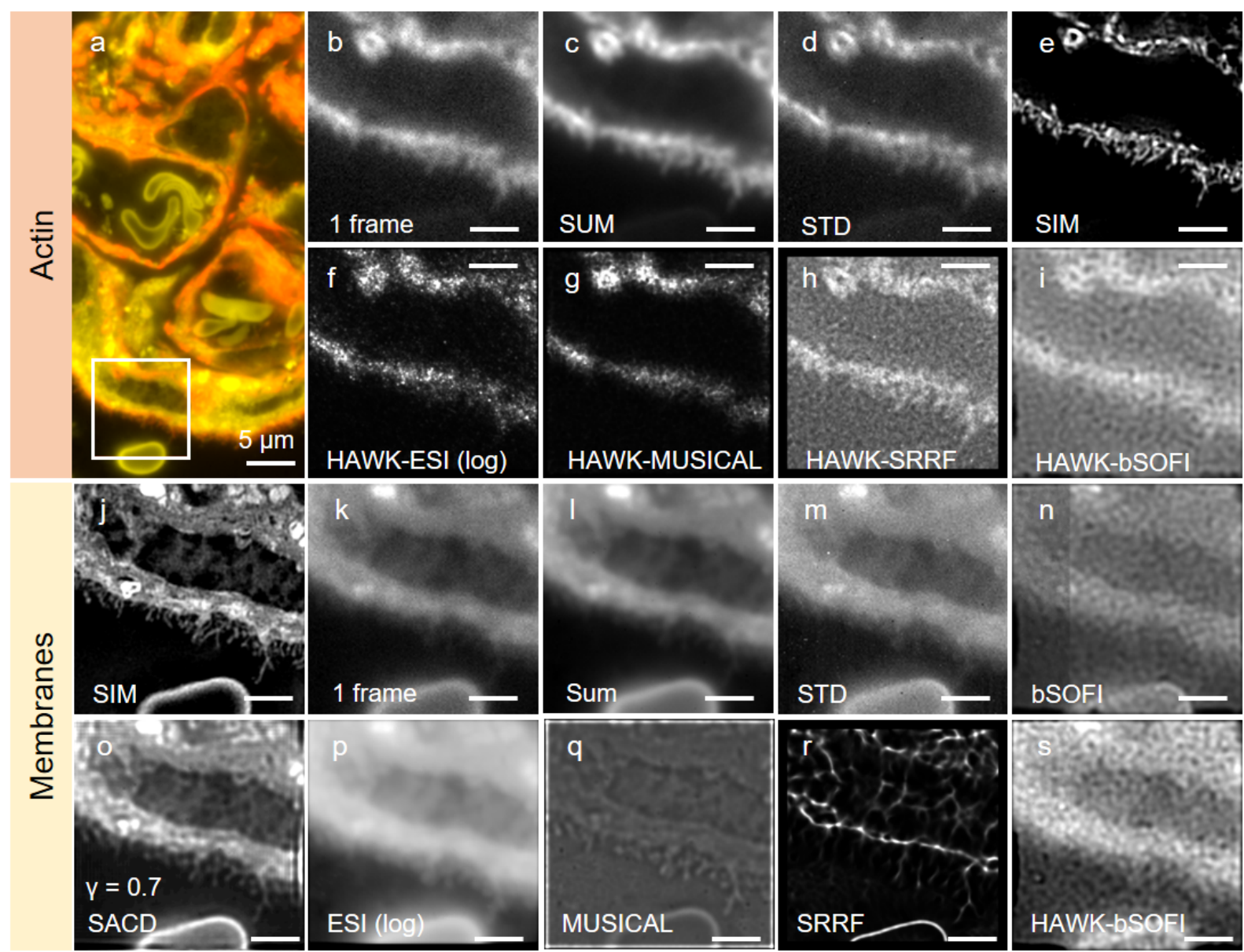

**Figure S10**

*Placenta tissue*: epi-fluorescent images of 1 µm-thick human placenta cryo-section labelled with phalloidin-ATTO647N and CellMask Orange for identification of F-actin and lipid membranes, respectively. (a) Overlaid sum image provides a large FOV of the two analyzed channels. F-actin displayed in red and membranes displayed in yellow; (b-s) Magnified view of the outlined area in *a*; (b-d) Overview of the sample data for the F-actin channel; (b) first frame of the image stack; (c) sum image of the image stack (400 frames); (d) standard deviation image of the image stack; (e) SIM reference image of the F-actin channel; (f-i) Results of combining HAWK (using 5 levels) along with different fluctuation methods. For this dataset, HAWK did not improve the reconstructions compared to the previous results (see Figure 5 in the main manuscript). The two bottom rows display an overview of the sample data and results for the membrane channel; (j) SIM reference image of the membrane channel; (k) first frame of the image stack; (l) sum image of the image stack; (m) standard deviation image of the image stack; (n-s) For the membrane dataset, the reconstruction algorithms exhibit similar performance as with the F-actin dataset, where SACD and MUSICAL performed the best among the techniques, revealing structural details of the tissue samples beyond what is visible in the sum image and in accordance with the SIM reference image; (i, s) Despite its potential of improving bSOFI reconstruction, the combination HAWK-bSOFI yielded degraded results for these datasets. 400 frames were used for the reconstructions. The scale bars are 2 µm.

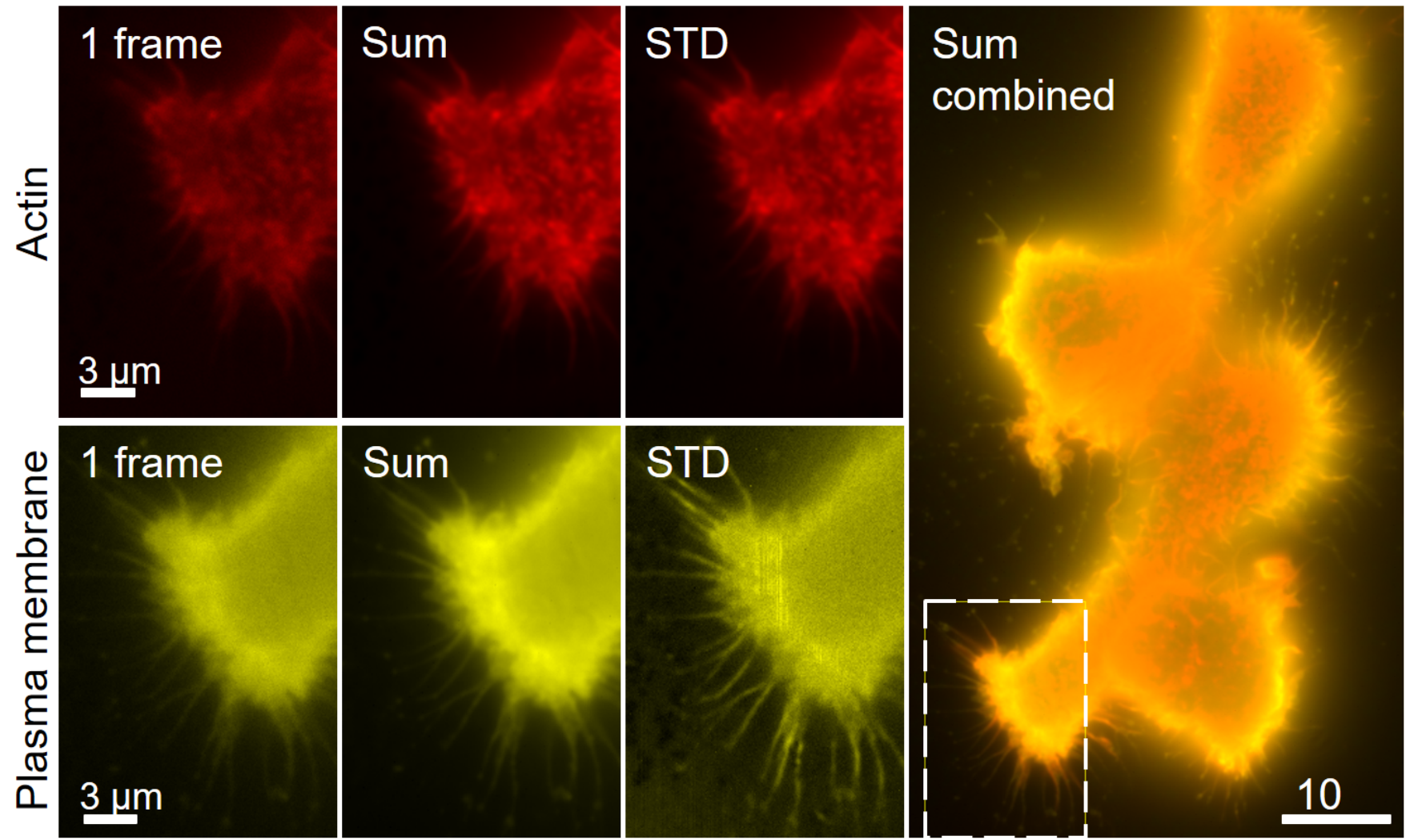

**Figure S11**

*Overview of the (epi) dataset used for the study of fixed-cell.* The panels show two-channel epi-fluorescence microscopy images of cultured macrophages labelled for F-actin (phalloidin-ATTO-647N) and the plasma membrane (CellMask Orange). The panels show a single image, the sum and the standard deviation image (STD) of 500 frames, together with a larger sample area with the smaller region subjected to analysis indicated. The reconstruction results are found in Figures S12 to S15.

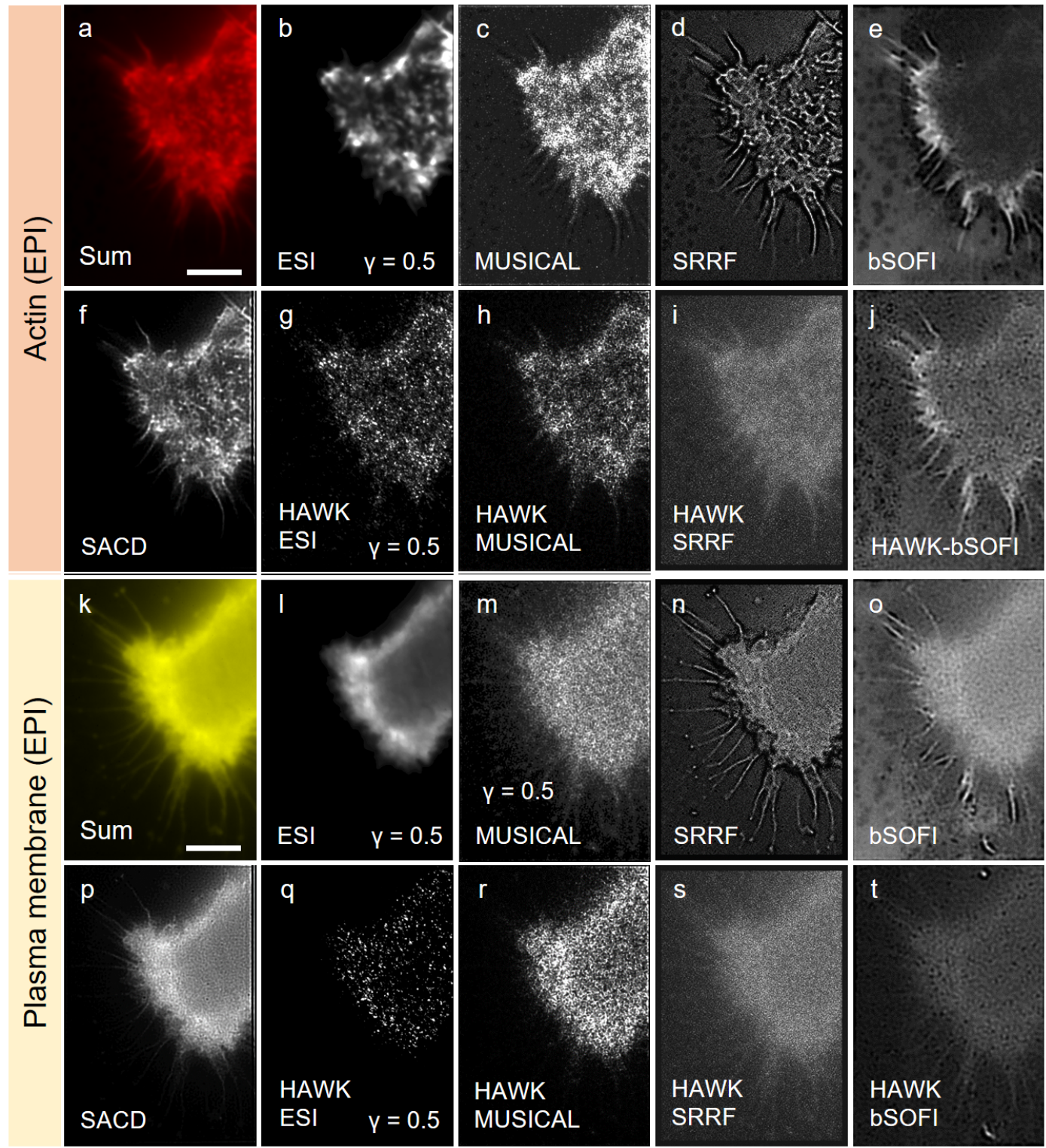

**Figure S12**

*Results summary of fixed-cell (epi):* Comparison of (the best) results for the different methods on fixed macrophages using epifluorescence microscopy. (a) Sum of 500 frames of the raw data stack (F-actin with phalloidin-ATTO-647N). The scale bar is 5 µm and apply to all panels; (b) ESI (order 4) on 500 frames. The image is γ = 0.5 intensity adjusted; (c) MUSICAL on 50 frames using threshold -1.0; (d) SRRF result on 50 frames using option TRM and radius 0.5; (e) bSOFI using 500 frames; (f) SACD with MPAC order 2 and 16 frames; (g) HAWK ESI on 4886 frames (resulting from 5 level HAWK on 500 raw frames). The image is γ = 0.5 intensity adjusted; (h) HAWK MUSICAL on 386 frames (resulting from 5 level HAWK on 50 raw frames) using threshold -0.3; (i) HAWK SRRF using 50 frames with the TRM option and radius 0.5; (j) HAWK bSOFI on 4886 frames (resulting from 5 level HAWK on 500 raw frames); (k) Sum of 500 frames of the raw data stack (plasma membrane using CellMask Orange). The scale bar is 5 µm; (l) ESI (order 4) on 500 frames. The image is γ = 0.5 intensity adjusted; (m) MUSICAL on 50 frames using threshold -1.4; (n) SRRF result on 50 frames using option TRM and radius 0.5; (o) bSOFI using 500 frames; (p) SACD with MPAC order 2 and 16 frames; (q) HAWK ESI on 4886 frames (resulting from 5 level HAWK on 500 raw frames). The image is γ = 0.5 intensity adjusted; (r) HAWK MUSICAL using 386 frames (resulting from 5 level HAWK on 50 raw frames) and threshold -0.5; (s) HAWK SRRF using 50 frames with the TRM option and radius 0.5; (t) HAWK bSOFI on 4886 frames (resulting from 5 level HAWK).

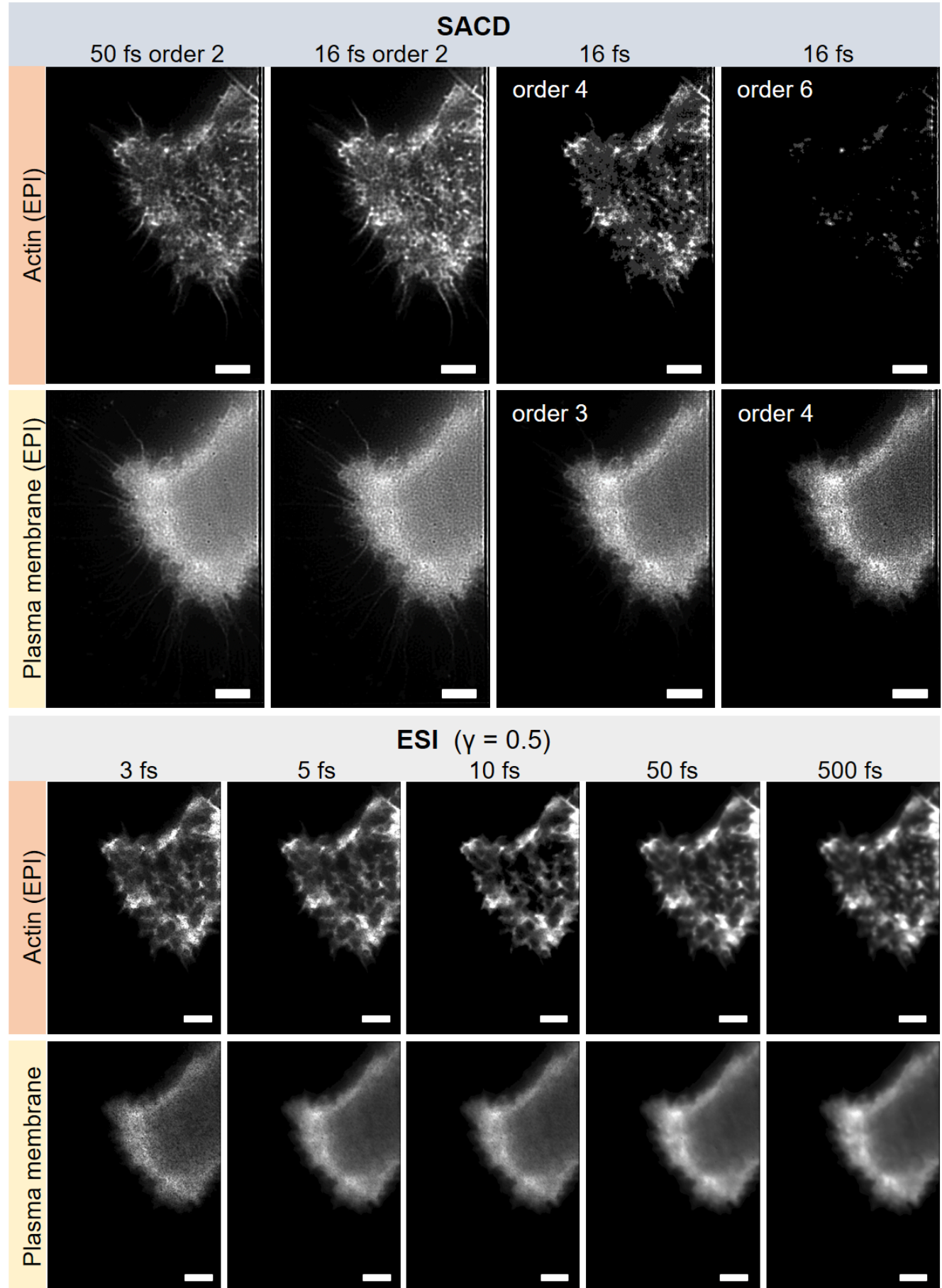

**Figure S13**

*Fixed cell results of SACD and ESI* using additional options/parameter as indicated in the figure. **SACD**: As also seen for the simulated data, increasing the number of frames beyond 16 does not significantly improve or change the SACD results. Increasing the MPAC order causes gradual disappearance of cellular structures. **ESI:** The ESI results do not appear to improve by increasing the number of frames beyond 5 frames. The results look similar whether 5 or 500 frames are used. To enhance the visibility of finer details, the ESI images are γ = 0.5 intensity adjusted. The scale bars are 2 μm.

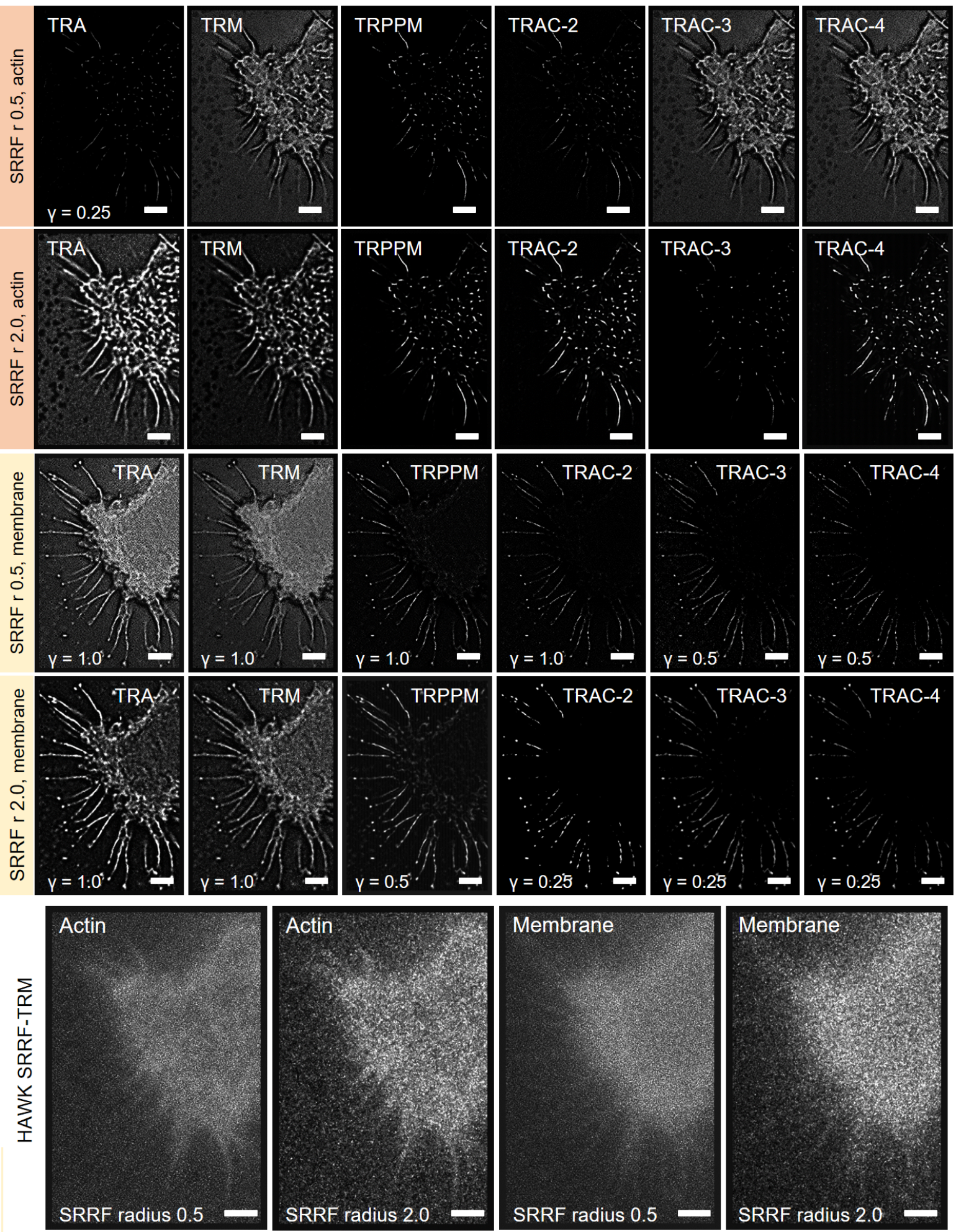

**Figure S14**

*Fixed-cell (epi): SRRF* results using additional options/parameter as indicated in the figure, all with 50 frames used for the image reconstruction. The scale bars are 2 μm. The many different options offered by SRRF are problematic as they can generate very different pictures and cellular structures, which leaves the users with difficult and subjective choices about which option, if any, accurately describes the (unknown) nanostructural details of the sample. In the case of SRRF, HAWK only produced image degradation for our samples.

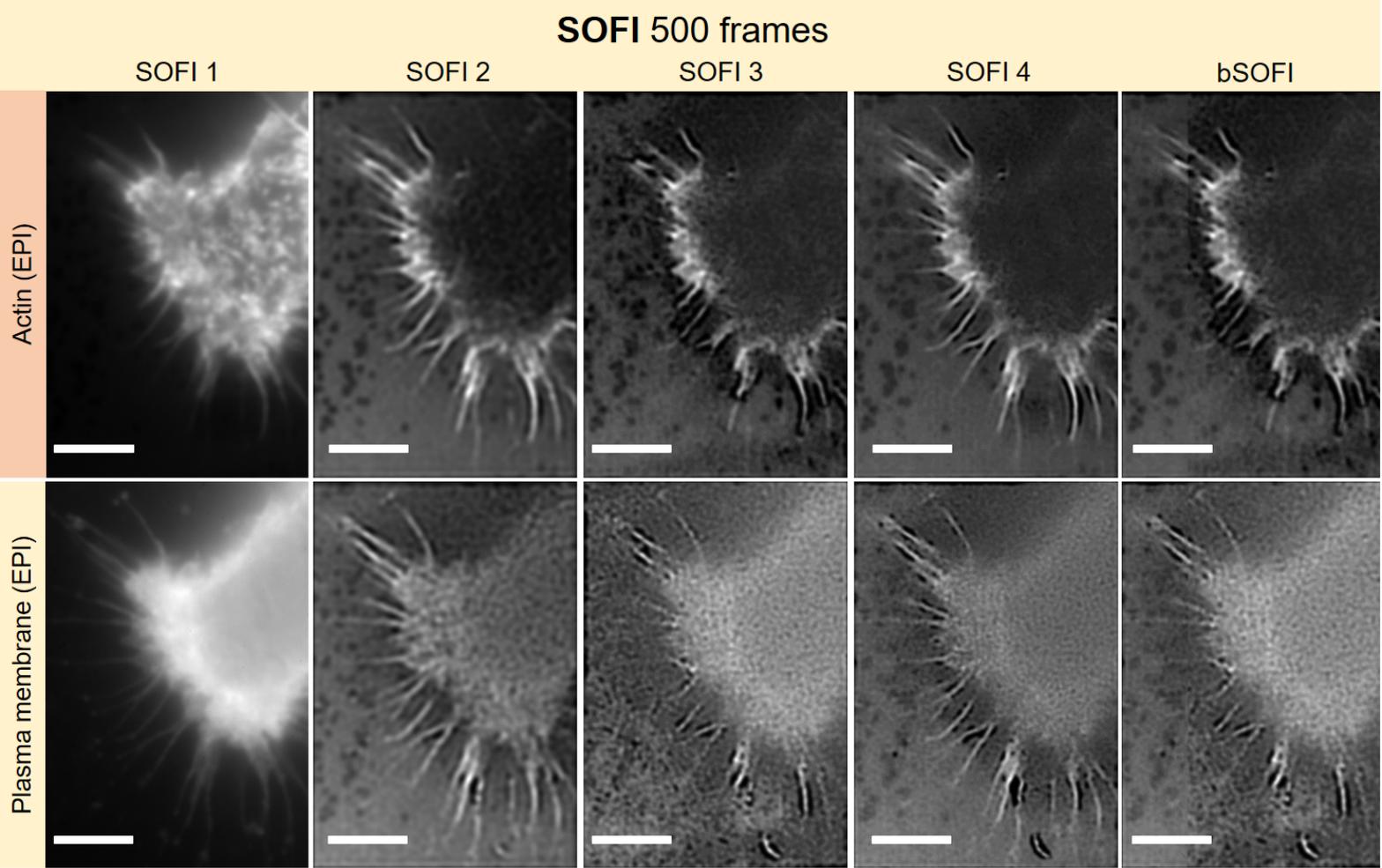

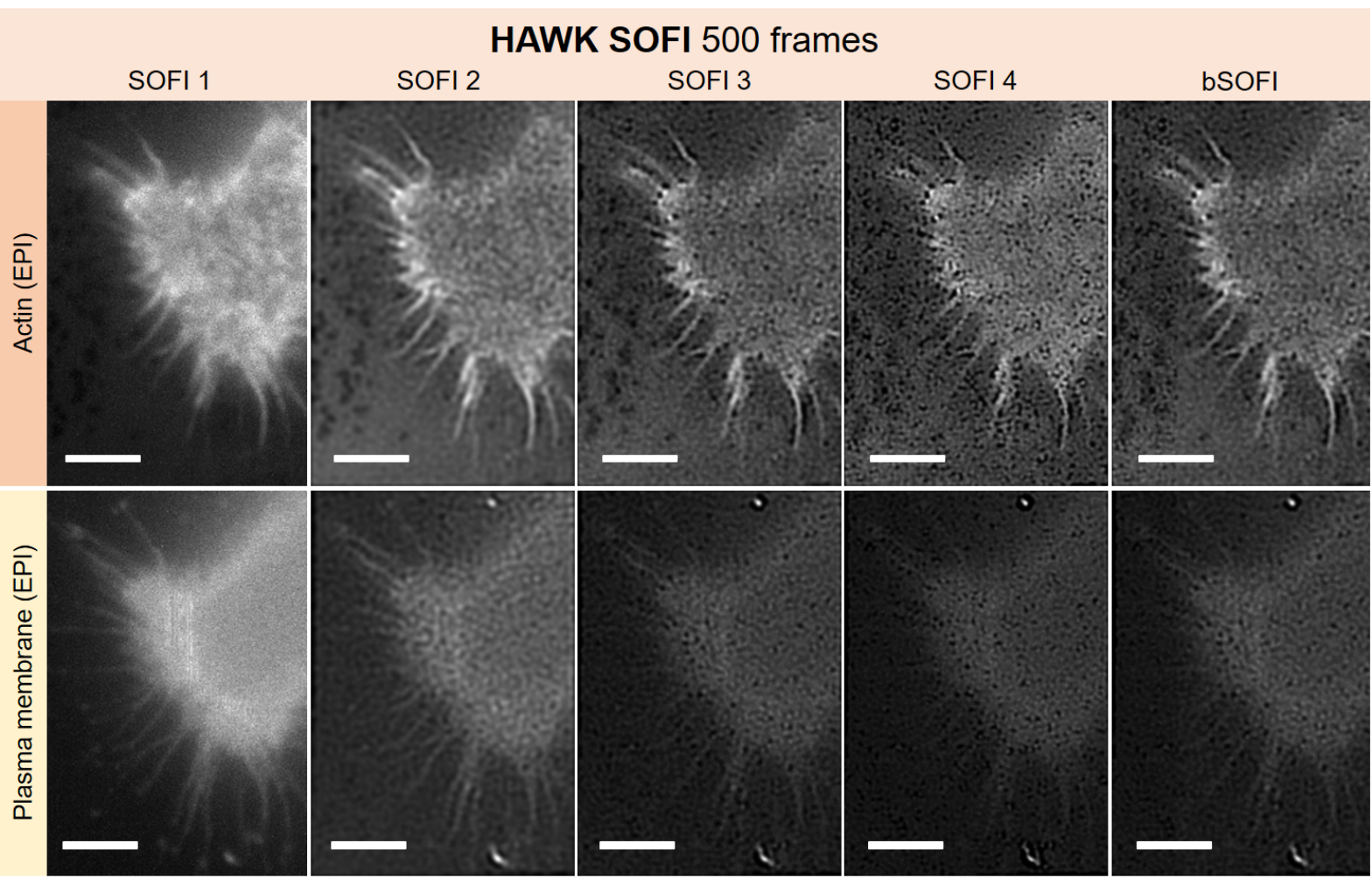

**Figure S15**

*Fixed-cell (epi):* SOFI and HAWK SOFI results using additional options/parameter as indicated in the figure. The results appear not to accurately describe the nanoscopic details of actin of membrane in macrophages. As indicated by simulations, this can be explained by low signal fluctuation of these fluorescence labels and also possible 500 frames not being sufficient for reliable SOFI reconstruction. The scale bars are 5 μm.

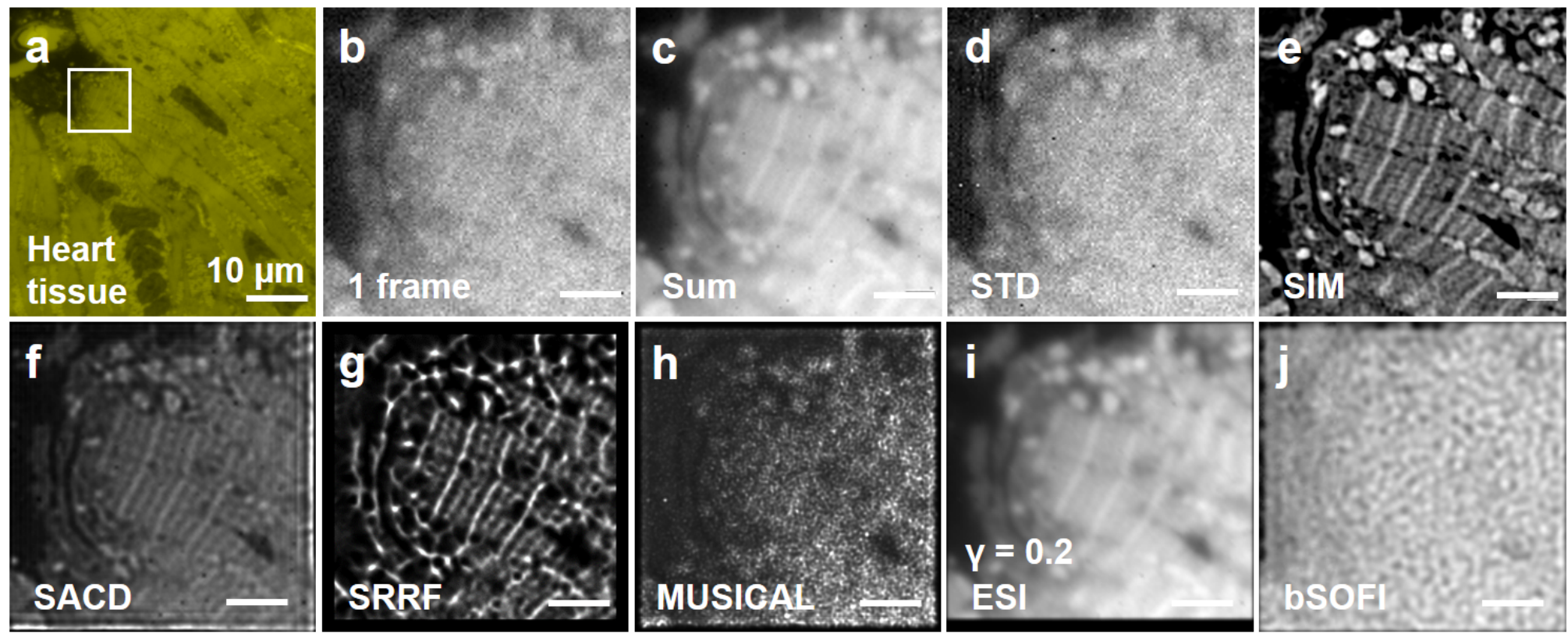

**Figure S16**

*Heart tissue:* epi-fluorescent images of 100 nm-thick pig heart cryo-section labelled with CellMask Orange for identification of lipid membranes. (a) Large FOV image of the cardiac tissue; (b-j) Magnified view of the outlined area in *a.* The scale bars are 2 μm; (b) first frame of the epi-fluorescent stack. The image indicates a poor signal-to-noise ratio; (c) Sum image of the stack used for the reconstructions; (d) Standard deviation image of the same stack; (e) The reference structured illumination microscopy image, which indicates not just better contrast than the sum image but also preserving the low contrast striations throughout the selected region; (f) The SACD reconstruction shows the best correspondence with the SIM reference image; (g) The SRRF reconstruction displays subtle artifacts compared to the SIM reference image (compare e.g. larger bright spots (mitochondria) in the SIM image, and the black circles present only in the SRRF image); (h) The MUSICAL result here exhibits noticeable artifacts. (i) The ESI reconstruction (gamma transformed 0.2) is similar to the sum image in panel *c.* (j) The bSOFI result displays noticeable artifacts. This sample gives no out-of-focus signal, but with a high noise level and high labeling density (therefore a low level of fluctuations).

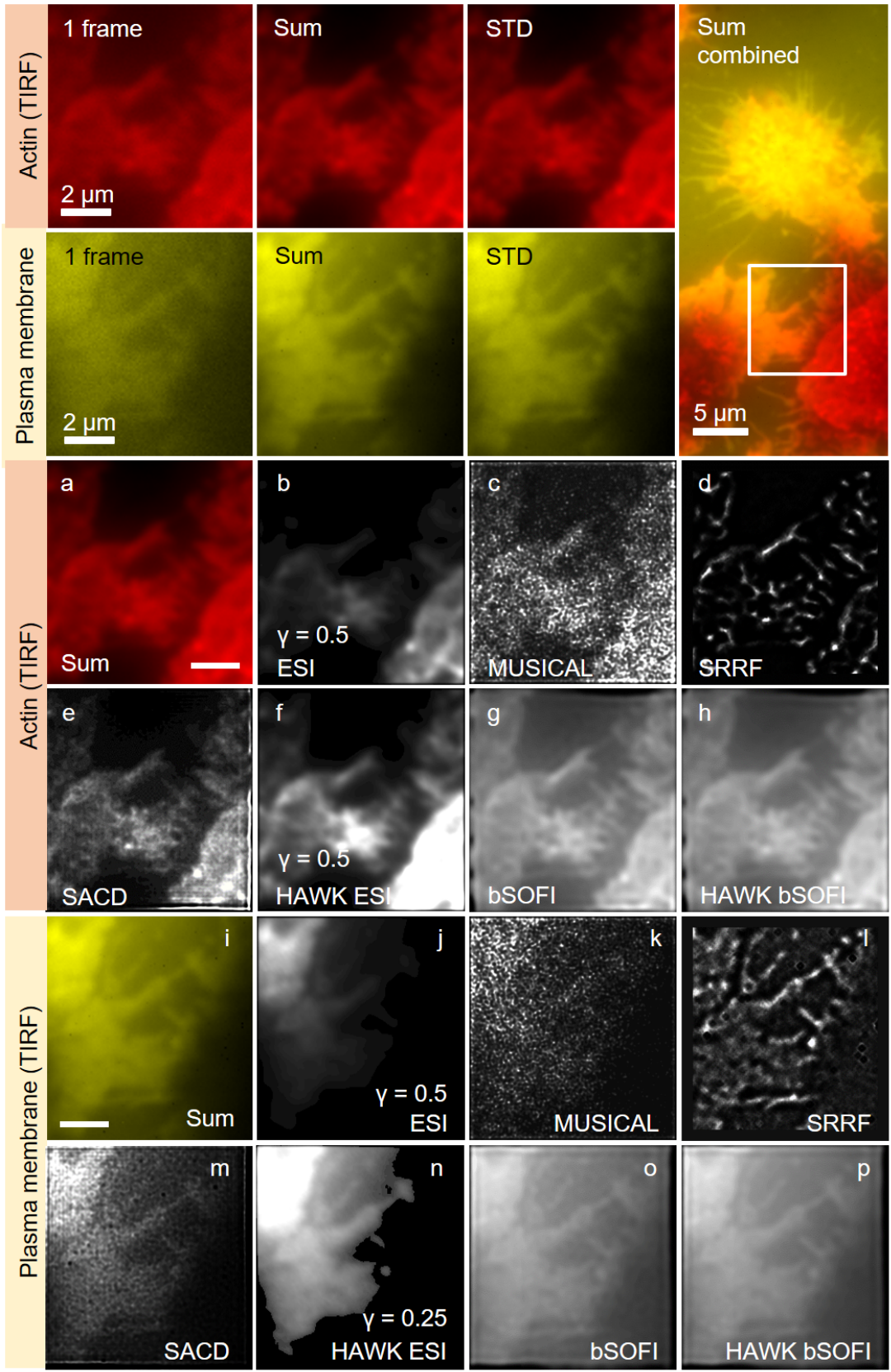

**Figure S17**

*Overview of the TIRFM datasets (top two rows) and results summary for fixed-cells.* The panels show F-actin (phalloidin-ATTO-647N) and the plasma membrane (CellMask Orange) of fixed macrophages. The panels show a single image, the sum and the standard deviation image (STD) of 500 frames, together with a larger sample area with the smaller region subjected to analysis indicated. The signal from CellMask dye caught by the coverslip surface became dominating while using TIRF illumination. *Panels a-p:* Comparison of (the best) results for the different methods on fixed macrophages using TIRFM. (a) Sum of 500 frames of the raw data stack (F-actin with phalloidin-ATTO-647N). The scale bar is 2 µm and apply to all panels; (b) ESI (order 4) on 500 frames. The image is γ = 0.5 intensity adjusted; (c) MUSICAL on 50 frames using threshold -1.3; (d) SRRF result on 50 frames using option TRM and radius 2.0; (e) SACD using MPAC order 3 and 16 frames; (f) HAWK ESI (5 level HAWK resulting in 4886 frames). The image is γ = 0.5 intensity adjusted; (g) bSOFI using 500 frames; (h) HAWK bSOFI on 4886 frames (resulting from 5 level HAWK). (i) Sum of 500 frames of the raw data stack (plasma membrane using CellMask Orange). (j) ESI (order 4) on 500 frames. The image is γ = 0.5 intensity adjusted; (k) MUSICAL on 50 frames using threshold -1.3; (l) SRRF result on 50 frames using option TRM and radius 2.0; (m) SACD using MPAC order 3 and 16 frames; (n) HAWK ESI (5 level HAWK resulting in 4886 frames). The image is γ = 0.5 intensity adjusted; (o) bSOFI using 500 frames; (p) HAWK bSOFI on 4886 frames (resulting from 5 level HAWK).

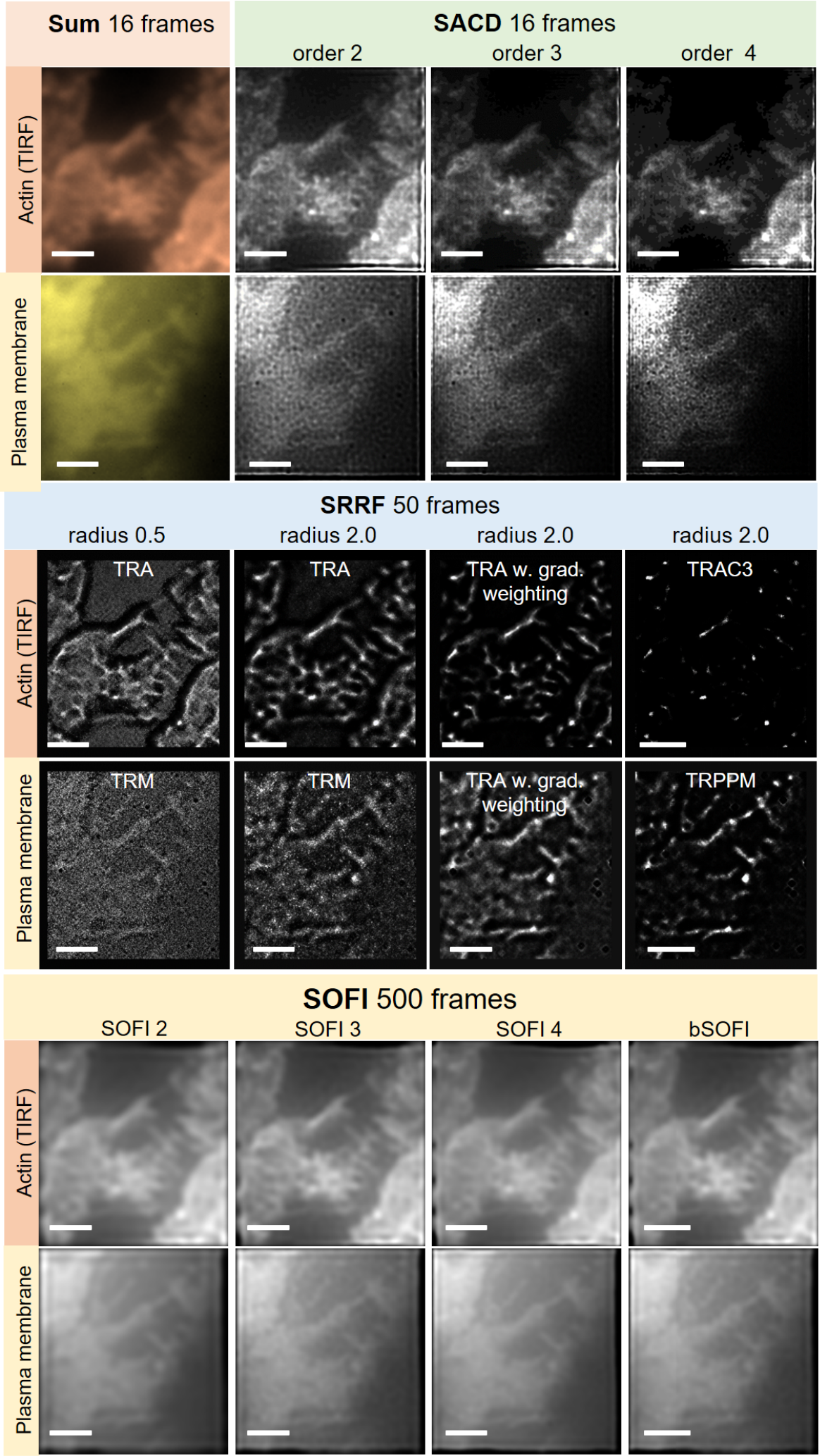

**Figure S18**

*TIRFM of fixed macrophages:* results using additional options/ parameters (indicated in the figure) for SACD, SRRF and SOFI.

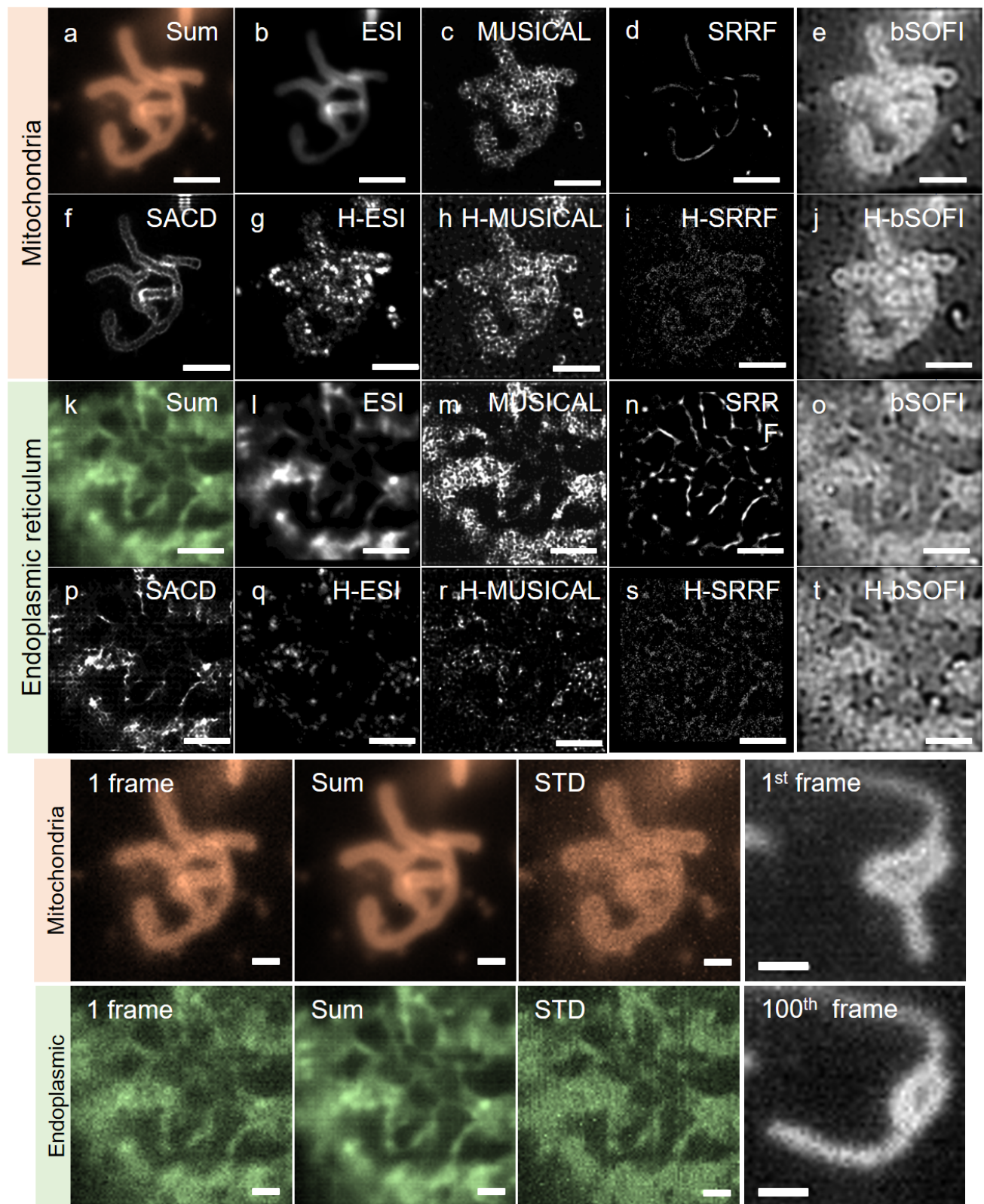

**Figure S19**

*Results summary and data overview for live-cell data:* Comparison of (the best) results for the different methods on live-cell image data (cardiomyoblasts cell-line H9c2) acquired using epifluorescence microscopy. Panels (a)-(j) are of mitochondria (outer membrane protein 25, OMP25-mCherry), and panels (k)-(t) are of the endoplasmic reticulum (KDEL-EGFP). The two datasets are two different color channels from the same cell, region and time-lapse sequence.

(a) Sum of the 64 frames (of mitochondria) used for the image reconstructions of panels (a)-(j); (b) ESI order 4; (c) MUSICAL using threshold -0.8; (d) SRRF using option TRAC order 2 and radius 0.5; (f) SACD using MPAC order 4; (g) HAWK ESI (5 level HAWK resulting in 526 frames). The image is γ = 0.5 intensity adjusted; (h) HAWK MUSICAL on 526 frames (resulting from 5 level HAWK) using threshold 0.0; (i) HAWK SRRF using option TRAC order 2 and radius 0.5; (j) HAWK bSOFI on 526 frames (resulting from 5 level HAWK);

(k) Sum of the 64 frames (of the endoplasmic reticulum) used for the image reconstructions of panels (l)-(t); (l) ESI order 4; (m) MUSICAL using threshold -0.5; (n) SRRF using option TRAC order 2 and radius 0.5; (p) SACD using MPAC order 4; (q) HAWK ESI (5 level HAWK resulting in 526 frames). The image is γ = 0.5 intensity adjusted; (r) HAWK MUSICAL on 526 frames (resulting from 5 level HAWK) using threshold 0.2; (s) HAWK SRRF using option TRAC order 2 and radius 0.5; (t) HAWK bSOFI on 526 frames (resulting from 5 level HAWK). The scale bars are 2 μm.

*Bottom rows, data overview*: 1st frame, sum and standard deviation (STD) image of the datasets used for panels a-t. 1st and 100th (last) frame of the 100-frame image stack used for the reconstructions of panels (g)-(l) of Figure 6 of the main manuscript. The scale bars are 1 μm.